\documentstyle[referee]{mn}
\def\baselinestretch{1}

\newcommand{\ci}{\'{\i}}

\newcommand{\be}{\begin{equation}}
\newcommand{\ee}{\end{equation}}
\newcommand{\eti}{ et al. \/}

\input psfig.sty

\begin{document}

\title{ Dark Matter Halo Structure in CDM Hydrodynamical Simulations}
\author[P.B. Tissera, R. Dom\ci nguez-Tenreiro ]
{P.B.Tissera $^{1}$ $^{2}$,
 R. Dom\ci nguez-Tenreiro $^{1}$ \\
 $^{1}$Postal Address: Departamento de F\'{\i}sica Te\'orica C-XI, Universidad
Aut\'onoma de Madrid, Cantoblanco 28049, Madrid, Spain\\
$^2$ Imperial College of Science, Technology and Medicine,
Blackett Laboratory, Prince Consort Road, London, United Kingdom\\
E-mail:rosa@astrohp.ft.uam.es
}

\maketitle
\def\baselinestretch{1}
\begin{abstract}

We have
 carried out a comparative analysis of the properties of dark matter halos
in N-body and hydrodynamical simulations. We analyze their density profiles,
shapes and kinematical properties with the aim of assessing the effects
that hydrodynamical processes might produce on the evolution of
the dark matter component.
The simulations performed allow us to reproduce dark matter halos with
 high resolution, although the range of circular velocities is limited.
We find that for halos with
circular velocities of   $[150-200] \ \rm km  \ s^{-1}$
at the virial radius, the presence of
 baryons affects the evolution of the
dark matter component in the central region modifying  the density
profiles, shapes and velocity dispersions.
We also analyze the rotation velocity curves of
disk-like structures and compare them with observational results.

\end{abstract}

\def\baselinestretch{1.5}

\vspace{2.in}
\pagebreak
\section{Introduction}
The importance of understanding the formation and evolution of dark matter
halos is based upon the
hypothesis that they contain relevant information about cosmological parameters
and the power spectrum of initial density fluctuations.
Theoretical works based mainly on the second infall model, first studied by
Gunn \& Gott (1972), have predicted different behavior for the density profiles
of dark matter halos.  These models simplified the problem of
the collapse of the structure by assuming spherical symmetry and
purely radial motions. Fillmore \& Goldreich (1984) and Bertschinger (1985)
found self-similarity solutions
 for the secondary infall collapse. These models predict
the formation of virialized objects with dark matter density profiles that are
approximately isothermal. Moutarde \eti (1995) show that this result is not
restricted  to the fully spherically symmetric case.
Hoffman \& Shaham (1985) and Hoffman (1988) extended the analysis to models
with different density parameters ($\Omega$) and power spectra ($P(k) \propto
k^{n}, -3<n<4$) concluding that density profiles of
dark matter halos should steepen for larger values of $n$ and lower $\Omega$.
However, because of the oversimplified hypothesis assumed, these models fail to
describe consistently the collapse and evolution of the structure
in a hierarchical clustering scenario where the structure is formed through
the collapse and aggregation of substructure. Numerical simulations
are well suited to follow the hierarchical built-up of halos througthout their
linear and non-linear regime (Frenk \eti 1985, 1988, Quinn \eti 1986, Efstathiou \eti 1988,
Zurek \eti 1988, Dubinski \& Carlberg 1991, Warren \eti 1992).
During the last years there have been  numerous works on the properties of
dark matter halos and their implications for explaining some observational
facts such as the rotation velocity curves of disk galaxies, cores of
dwarf galaxies, gravitational lenses of background galaxies by galaxy  clusters, etc.
These techniques have resulted in a very useful tool limited mainly by
numerical resolution. Several works have been carried out recently
 which have taken  advantage of the last improvements in
computational facilites and numerical techniques in order
to perform high resolution numerical simulations.

In particular, several works have been dedicated to the study
of dark halo density profiles in N-body numerical simulations where only
the gravitational field is integrated.
The most recent works by Navarro \eti (1995, hereafter NFW, 1996a,b ),
 Cole \& Lacey (1997), Tormen \eti (1997),
 have shown that
halos in purely dynamical simulations in scale-free and Cold Dark Matter (CDM)
universes in low-density, flat and open cosmogonies, can be fitted by the following curve:

\be
\rho(r)\propto \frac{x^{-\alpha}}{(1+x^{\beta})^{\gamma}}
\ee

where $x=r/r_{s}$, with a single scale radius $r_{s}$.
Hernquist (1990, hereafter Hern) proposed the combination $(\alpha,
\beta, \gamma)=(1,1,3)$ for
spherical galaxies and bulges of disk galaxies,
 while NFW
fitted equation 1 with $(\alpha,\beta,\gamma)=(1,1,2)$.
Tormen \eti (1997) analyzed the evolution of cluster halos
in scale-free universes in high resolution numerical simulations, and compared
the fits given by the models proposed by Hern and NFW.
They found that both models described the simulated density profiles with
good accuracy.

Hence, in general terms, all works seem to agree on the fact that halo density
profiles in N-body simulations of scale-free and CDM models in open or flat cosmogonies
 are curved and
well described by equation 1. According to the definition proposed by NFW,
this profile has  two free parameters:
$c$, the concentration parameter   and $\delta_{c}$, the
 characteristic density.
They  seem to depend on
the initial conditions and the mass of the halos. The characteristic density,
$\delta_{c}$, is
found to be proportional to the density of the universe at the time of
collapse of the objects which merge to form the core halo.
These authors have pointed out that dark matter density profiles are always well-described
by equation (1) and that only the values of its parameters show some dependence with the spectrum or
cosmology. In particular, it has been asserted in their works that the only primordial
information that dark matter profiles remember is their time of collapse.

So far all recent works have dealt with  dark matter halos
in purely gravitational scenarios,
 describing its evolution and properties as
a function of mass and initial conditions.
However, halos contain a fraction of baryonic  matter that,
although  in a smaller proportion, may influence the evolution of the
dark matter.
There have been several attempts to address the effects of dissipation
on the evolution of the dark matter component in galactic halos
using different approaches
to model it (Blumenthal \eti 1986, Flores \eti 1993, Flores \& Primack 1994,
 Dubinski 1994).
Blumenthal \eti (1986) used analytic models and dissipative particle collisions
in N-body simulations to study the response of the dark matter halo to
the presence of baryons, with the aim at understanding the origin of rotation
curves of spiral galaxies. Subsequent works (Flores \eti 1993, Flores \&
Primack 1994)
used their results to deep on the analysis of dark matter structure in halos.
Dubinski (1994) analyzed the effects of dissipation on the dark matter
halo by introducing
an analytical potential well representing disk-like and spheroidal structures
in its central region. This author found that
when the baryonic component is taken into account, the overall potential
well is modified in the central region leading to a change in the density
profiles and
shapes, in agreement with the result of Blumenthal \eti (1986).
Evrard \eti (1994) analyzed the formation of galaxies in cosmological
 hydrodynamical simulations,
finding that halos are more spherical  than in purely dynamical
ones.

Regarding clusters of galaxies, there have been several numerical works
on the analysis of their density profiles. Observational estimates of gas
and dark matter density profiles in X-ray clusters show controversial results
since gas density profiles seem to be shallower ($\rho\propto r^{-1}$)
than  those of the dark matter component, deduced by gravitational lensing techniques.
Navarro \eti (1995) analyzed hydrodynamical simulations of
galactic clusters and found that they could be fitted  by a shallower power law:
$\rho\propto r^{-1}$, in the central region. On the other hand, Aninos \& Norman (1996)
obtained a different result from high resolution simulations of galaxy
clusters using a two-level grid code. They find that both gas and dark
matter density profiles follow a unique power law $\rho \propto r^{-9/4}$.
They pointed out the presence of a non-convergent gas density core even
for their highest resolution run, in disagreement with Navarro \eti (1995) results.
 It is an uncomfortable situation
that numerical studies of  dark matter halos of cluster of galaxies using
different techniques yield different results.
Numerical resolution problems and artefacts are still unavoidable limitations,
together with our ignorance of the detailed physics involved in the formation
of the structure in the Universe.

From an analytical point of view, Chieze \eti (1997) and Teyssier \eti (1997)
studied the adiabatic
collapse of spherical symmetric perturbations of gas and dark matter. They
found a segregation between gas and dark matter components. The dark matter
density profiles  was found to be steeper than that of the gas. According
to these authors, the main effect of the joint evolution of their adiabatic
purely spherical collapse is the compression of the gas core by the dark matter.
But they found no changes in the dark matter density profiles.
Their hypothesis of geometry and non-treatment of cooling effects may
result in differences in the final properties of the gas and dark matter
distributions when compared to the results of  cosmological hydrodynamical
 simulations which also include cooling mechanisms.

Studying halos which contain both  components, baryonic and dark matter,
with numerical techniques is still a  quite
difficult task to accomplish, principally because of the high resolution
required by the dissipative component, but also because of the  complexity
of the different processes related to the physics of baryons  such as star
 formation, supernovae, etc.
The aim of this paper is to present a comparative analysis of galactic
halos formed
in simulations with and without hydrodynamical effects, and to assess
 any changes
in the properties of the dark matter arising as a consequence of the presence
of the dissipative component.
Our results  attempt to describe the effects of
the joint evolution of the  dark matter
and baryons, principally in the central region of galactic-like halos.
They do not intend to be conclusive because of our numerical
resolution limits. We  also attempt to estimate
the effects of star formation history on the properties of the objects
at $z=0$.
This paper is organized as follows. Section 2 describes the simulations.
Section 3 deals with the analysis of dark matter halos and disk-like objects
properties, and discusses the results. Section 4
outlines the conclusions.

\section{Simulations}
The initial distributions of positions and
velocities of particles in all simulations
have been set using COSMICS ( kindly provided by E. Bertschinger ). They are consistent with standard
CDM cosmologies: $\Omega =1, \Omega_{b}=0.1, \Lambda=0$, $b\equiv 1/\sigma_{8}= 2.5$ or $1.7$
where $\Omega$ and  $\Omega_{b}$ are the cosmological and
the baryonic density parameter, respectively,
 $\Lambda$ is the cosmological constant
and $b$, the bias parameter. We have adopted a Hubble constant of
$H_{o}=100 h \ \rm{km s^{-1} Mpc^{-1}}$ with $h=0.5$.
 We studied two sets of simulations with different initial conditions (set I and II). The
simulations in each set share the same initial conditions but have different
hydrodynamical parameters.
In all simulations we followed the evolution of 262144 particles in a
 periodic box
 of 10 Mpc.
 All dark, gas and star particles have the
same mass, $2.6\times10^{8} \ M_{\odot}$. The initial distribution of
gas particles has been choosen randomly from the total particle distribution
provided by COSMICS.
The hydrodynamical ones (simulations I.2, I.3 and II.2)
 were performed using a Smooth Particle Hydrodynamical  (SPH) code (see
Tissera \eti 1997 for
details of the SPH implementation)
 coupled to the high resolution adaptative particle-particle-particle-mesh code
(AP3M), kindly provided by Thomas \& Couchman (1992) (simulations I.2 and
I.3 correponds to simulations 6 and 3 in Tissera \eti 1997, Table I).
For comparison, the same initial conditions were run taking into account
only gravitational effects  using the  AP3M (simulations I.1 and II.1).
The gravitational softening used was $\epsilon_g = 5$ kpc.
Unfortunately, this SPH code
does not allow us to use individual time steps.  The integrations were
carried out with 1000 steps of $\Delta t =1.3 \times 10^{7}{\rm yrs}$.
All simulations were integrated only gravitationally from $z>50$ to
 $z=10$, and  from there
the SPH forces were also taken into account (Evrard \eti 1994) up to $z = 0$.
Table I lists their main characteristics.

Hydrodynamical runs also include a star formation algorithm (Tissera \eti 1997)
which allows to transform cold gas in high density regions into stars.
The parameters $\eta$
 and $\rho_{star}$ in Table I
 are related to the star formation scheme: $\eta$ is the
 star formation efficiency and $\rho_{star}$ is a critical
 density above which
 a cold gas particle in a convergent flow is transformed into  stars.
The value of $\rho_{star}$ is estimated by requiring the cooling time
of the gas cloud represented by a particle to be smaller than its
dynamical time. The actual value of $\rho_{star}\approx
 7\times 10^{-26}{\rm g/cm^{3}}$  used here is a
minimun limit for the gas density to fulfill this requirement (Navarro $\&$
White 1993).
Because of the  high efficiency of the radiative cooling and  the fact that
no heating sources have been included,
the star formation process   is very efficient (a lower cutoff of $T=10^{4}$ K has been
used for the temperatures). This leads
to an early transformation of the gas into stars, and prevents the
formation of disk-like structures. As a consequence,
 in simulations I.3 and II.2  all objects are spheroids.
In simulation I.2 a smaller value of $\eta$ was used.
 This  allowed the gas
to settle into  disk-like structures, although the star formation history
of these objects was artificially delayed to low redshift. Objects in set II
are dynamically more evolved than those in set I because of their
higher $\sigma_{8}$ value. By the same reason, objects in run II.2 
have suffered 
from an earlier transformation of the gas into stars than those in run I.3,
even if they have the same values of the $\eta $
 and $\rho_{star} $ parameters.

The simulations performed allow us to
study the properties of a set of  typically resembling galactic-like  objects
 with
very different formation history and of their dark matter halos.
 Some of them have undergone multiple mergers
(Tissera \eti  1996) while others have experienced a
more quiescent evolution. So differences on the properties of halos
which may arise as a result of a different evolutionary path may be assessed.
However, we do not have   a
large range of circular velocities to study the dependence of the results
on the mass.

\section{Analysis}

As mentioned in the introduction, there have been
several analytical and numerical works on the properties of the dark matter
structure. Most
of them were carried out using N-body simulations which provide
the correct gravitational evolution of the dark matter. Nevertheless,
serveral authors (eg. Blumenthal \eti 1986, Carlberg \eti 1986, Flores \&
Primack 1994, Dubinski 1994) have pointed out that the presence
 of baryonic matter might
affect the evolution of the dark component by modifying the potential
well in the central region, and producing a change in  their
shapes and  density profiles.
However, there have not been  exhaustive studies of halos in SPH simulations.
Hence, it is very interesting to carry out a comparative analysis of the
properties of dark halos in purely dynamical and hydrodynamical simulations.
We  run sets of
simulations sharing the same initial conditions but with and without
hydrodynamical  effects. A direct comparison between the shapes and density
profiles and an assessment of the possible  effects caused by
dissipational mechanisms  can be then carried  out.

In this work, we analyzed only the more massive objects whose halos 
and the galaxy-like objects they host (GLOs, see below) are
represented by a fairly large number of particles.
We have chosen halos 
in hydrodynamical runs which contain  more than 300
baryonic particles at their virial radius ($r_{200}$). 
 Hence we describe the properties of the dark
halos of 20 galaxy-like objects in hydrodynamical simulations
and 12 in purely gravitational ones.
Halos have been identified at their virial radius according
to the discussion of White \& Frenk (1991). The virial radius is
the radius  where the density contrast of the fluctuation reaches a value of
$\delta\rho/ \rho \approx 200$.
In the hydrodynamical case, baryons are included so the virial radius
corresponds to that of the total mass distribution (baryons and dark matter).
In simulations of set II, we have discarded some objects that belong
to groups and therefore, share a common dark halo.  However, in set I
we keep halo 7 because it hosts only two main baryonic clumps which assembled
at very low redshift.
The halo mass center is defined by an interative method which searches for
the mass center of concentric spheres of decreasing radius. The process
is halted when the number of particles in the sphere equals a certain
fixed number. This algorithm has been proved to converge quickly.
The baryonic clumps within $r_{200}$ of a halo are isolated using a
friends-of-friends algorithm with a linking length of the order of $10\%$
the mean particle separation. We will refer to them as galaxy-like objects.
It has to be noted that the center of mass of the halos may not
coincide with the center of mass of the principal galaxy it hosts. This is due to
the presence of satellites within its virial radius. The difference between
both mass centres may be of order of $3-6$ kpc.
Table II  summarizes the principal characteristics of halos and the GLOs
they host (spheroids (S), disks (D) and pairs (SP or DP)).

\subsection{Shapes}

In order to analyze the shapes of dark matter halos,  we follow a procedure
based upon the discussion
presented by Curir \eti (1993, and references therein).
 Basically, we calculated a sequence of
concentric ellipsoids containing fractions of the total mass of the objects.
We identify inside this region
a sequence of 10 concentric ellipsoids containing
fractions of 10 \% to 100 \% of the total mass:
\be
r=[x^{2}+\frac{y^{2}}{(b/a)^{2}}+\frac{z^{2}}{(c/a)^{2}}]^{1/2}
\ee

The semiaxes of the triaxial ellipoids ($a>b>c$)
are calculated  iteratively :

\be
\frac{b}{a}=(\frac{I_{22}}{I_{11}})^{1/2},  \frac{c}{a}=(\frac{I_{33}}{I_{11}})^{1/2}
\ee
where  $I_{11}> I_{22} >I_{33}$ are the eigenvalues of the tensor of inertia
($I_{jk}=\sum_{i} \frac{x_{i}^{j}x_{i}^{k}}{[r_{i}]^{2}}$).
In order to determine $a,b$ and $c$, an iterative cicle is required starting  with
$\frac{b}{a}=\frac{c}{a}=1$. The eigenvalues are calculated
for these first axial ratios, and then a new set of semiaxes is estimated
which constitutes the input of the next iteration. The procedure is
repeated until the axial ratios  $\frac{b}{a}$ and $ \frac{c}{a}$ have
a percentage change of less than $10^{-3}$. The axial ratios are estimated
for each of the ten ellipsoids.

\begin{figure*}
\centering \psfig{file=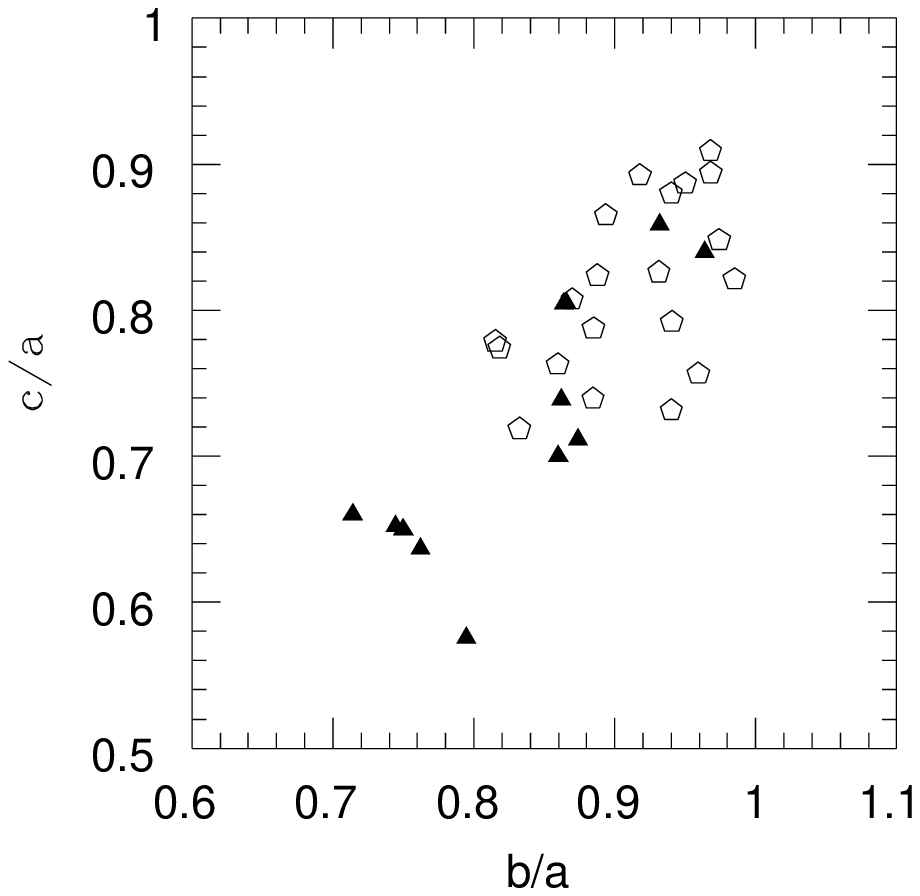,height=13cm,rheight=7cm,clips=,angle=0}
\caption{ }
\end{figure*}

Fig. 1 shows the ratios  $b/a$ and $c/a$ for all halos in simulations I.2, I.3
and II.2 (open pentagons),
 and their dissipationless counterparts in simulation I.1 and II.1
 (solid triangles) at the virial radius.
As already discussed by several authors (eg. Dubinski 1994, Evrard \eti 1994
), halos formed in purely
gravitational simulations tend to be more prolate. Halos in simulation
I.1 and II.1 have on average $<b/a>= 0.83 $, $<c/a>=0.78$
at $r_{200}$.
The same halos, but in the hydrodynamical runs, have modified their shapes and
are more spherical, and
they show, on average, a minor flattness
 $<b/a>\approx 0.91 $, $<c/a>\approx 0.81$.

Another estimator of shapes is  the parameter $T=(a^{2}-b^{2})/(a^{2}-c^{2})$.
For a purely prolate object $T=1$, while for a purely oblate,
 $T=0$.
The average value of $T$ for I.1 and II.1 is $<T>= 0.63$,
while for their  hydrodynamical counterparts,  $<T>= 0.51$.
This result shows how the shapes of halos change from  prolate to
 more oblate ones when the dissipational properties of the gas
are taken into account.

 An inspection of the different ellipsoids defined according to equation 2
shows the dependence of  shapes on  radius.
 From $r=100 $ kpc to $r_{200}$ the shapes are quite stable.
 So we
calculated the axis for a sequence of  10 ellipsoids
containing fractions from $10\%$ to $100\%$ of the dark mass within
$r=100$ kpc.
In Fig. 2a, we can see the ratios $b/a$ (open) and
$c/a$ (solid) for  halos containing a disk-like (pentagons)
and a spheroidal (triangles) structure (both objects are in simulation I.2).
 All objects tend to be more
prolate in the central region becoming more oblate-like with increasing radius.
If we compare the distribution of shapes of the halos with disks and
spheroids, we find that halos with disks tend to be slightly more
prolate in the central regions

\begin{figure*}
\centering \psfig{file=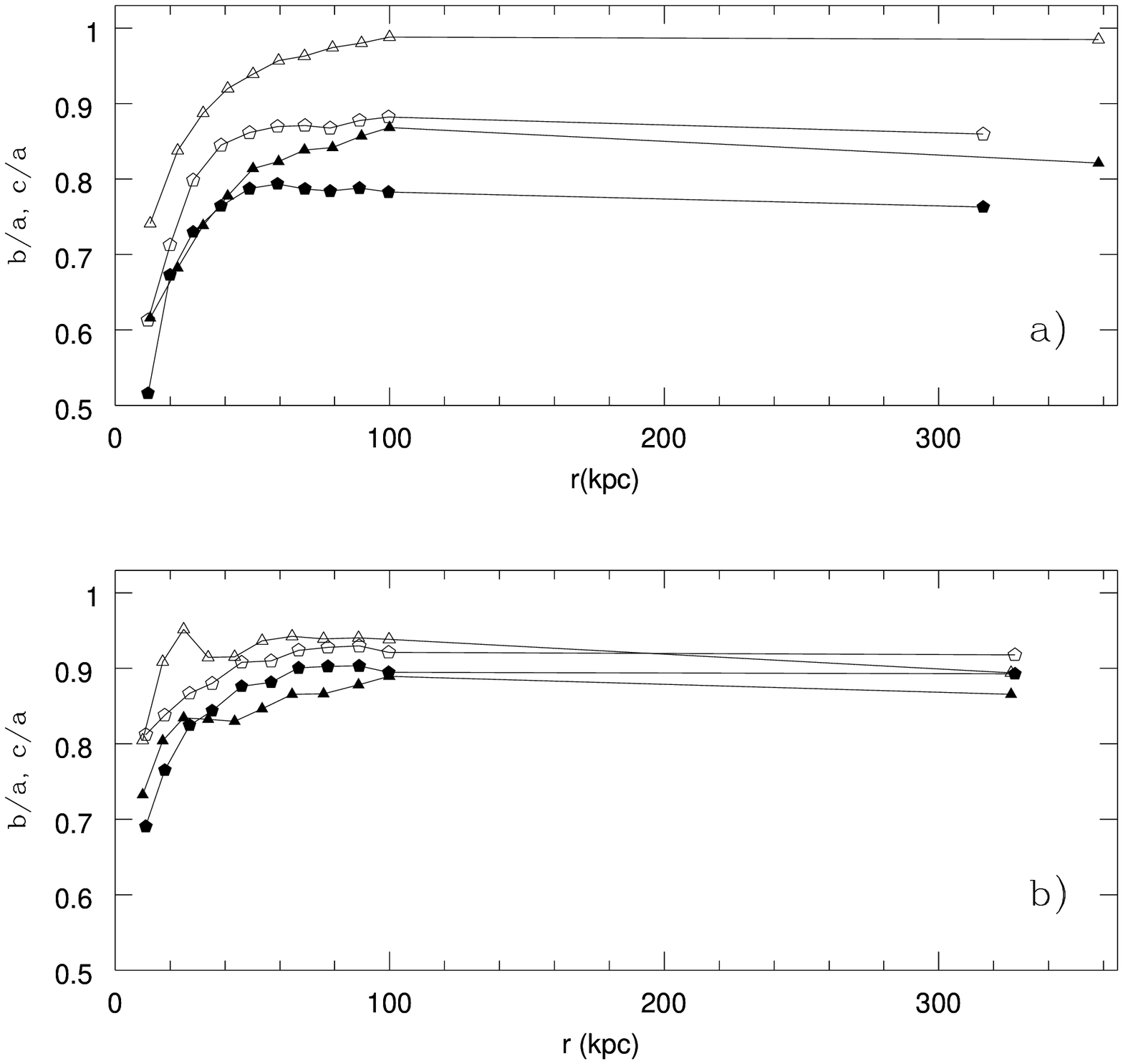,height=10cm,clips=,angle=0}
\caption{ }
\end{figure*}

In Fig. 2b, we compare the shapes of a halo containing a disk in I.2 (pentagons) and
its counterpart in I.3, where all objects are spheroids (triangles).
A weak trend to  more oblate structures is found in halos hosting spheroids.
 In this case, the ratio
 $c/a$ (solid triangles)
 is smaller than $c/a$  for halos with disks (solid pentagons), except in
the central region where the relation is the inverse.
 Although our sample
is quite small to draw a conclusion from it, the general trend agrees with
Dubinski's work who found that adding a spheroidal
 potential well at the center of a halo produced a stronger
effect on its shape than placing a disk-like potential.

\subsection{Kinematical Properties}

We calculated the velocity dispersion tensor in spherical coordinates in the
reference system of each halo:

\be
\sigma^{2}_{k}(r_{i})=\frac{1}{N+1} \sum_{j=i-N/2}^{i+N/2}\left[
V_{k}(r_{j})-\bar{V}_{k}(r_{i})\right]^{2}
\ee

where

\be
\bar{V}_{k}(r_{i})=\frac{1}{N+1}\sum _{j=i-N/2}^{i+N/2} V_{k}(r_{j})
\ee

and where $V_{k}$ (with $k=r, \theta,\phi$) are the velocity components
 in spherical coordinates, $N+1$ is the number of particles in a
bin (we have taken $N = 200$)
 and $r_{i}$ is the position of the $i^{th}$ dark matter
particle relative to the center
of mass.
  All velocity dispersions have been calculated at
  particle positions by first sorting the particles according to their
distance to the center of mass of the halo and, then,
 binning in concentric spherical shells of
a fixed $N+1$ number of particles, where each $i^{th}$ particle is placed
 at the middle point of the bin.

  In isotropic systems the velocity
dispersion in each coordinate should be $3^{-1/2}$ of the total dispersion
 ($\sigma$).
We found that $\sigma_{r} /\sigma$ is slightly greater than
$\sigma_{\theta} /\sigma$ and  $\sigma_{\phi} /\sigma$.
This fact is
reflected in the anisotropy parameter $\beta_{dark} = 1- \frac{\sigma_{\theta}^{2}}
{\sigma_{r}^{2}}$. Fig. 3a shows $\beta_{dark}$ for halo  4 in I.2 and halo 2 in I.3
(heavy lines),
 and
their gravitational counterparts (light lines) as examples.
 All halos are nearly isotropic at the center but
show an increasing anisotropy  as a function of the distance
to its mass center, reaching an average maximun of $\beta_{dark} \approx 0.3-0.5$.
So the outer regions of the halos are more dominated by radial dispersions.
There are no significant systematic differences
among the values of $\beta_{dark}$ of halos in different simulations.
However, small differences may
be covered up by the noise originated by the discreteness in the number of
particles.
In order to have an estimation of the strength of the signal over
this noise, we measured $\beta_{dark}$ for a simulated dark matter
halo of $10000$ particles consistent with a King model (King 1966). The rms
dispersion found was $\Delta rms \approx 0.11$.
 Hence, the anisotropy of the velocity distribution estimated
for the simulated halos can be regarded as a real signal.
On the other hand, the statistical significant
 features individualized in the graphics of
Fig.3a can be matched with the presence of satellites or substructures
within $r_{200}$. The positions of these substructures change within
simulations  mainly due to modifications in the collapse
 times of satellites
originated by hydrodynamical effects (Navarro \eti 1995).

By contrast, a difference is found among their total velocity dispersions.
 Fig. 3b shows the velocity dispersions of halos 1, 2, 3, 4
in simulations I.1 (light lines) and I.2 (heavy lines).
It can be seen from this figure that $\sigma$ of halos in
hydrodynamical simulations increases by a percentage of approximately
$ 59 \%$ between
$r=100$ kpc and $r=10$ kpc.
 In the purely gravitational simulations, however,
the percentage is significantly smaller, $ 10\%$,
remaining practically flat over the whole range of radii.
This effect is clearly related to the presence of a baryonic core in the
central region of the halo which modifies the overall potential well.


We have also compared the velocity dispersion profiles of halos
in simulations I.2 and I.3. As mentioned above, these simulations
differ from each other only on their star formation efficiency  which
has led to the formation of spheroidal objects in I.3, while,
conversely, I.2 has also disk-like
structures.
We found that the
velocity dispersion of a halo
hosting a disk and the same halo with a spheroid in its central region,
does not show any significant
difference in our simulations.

\subsection{Halo Density Profiles}

We  calculated the density profiles of dark matter halos for simulations
in sets I and II. The density profiles have  been defined by binning
the particles in spherical shells centred at the halo mass center.
We work with weighted profiles using shells containing  a fixed number of
particles ($30$ or $50$). We also estimated the density profiles
by binning in the logarithm of the radius. Both procedures yield the same
results. We have preferred to carry
out the analysis using the weighted profiles since the other one gives
too much weight to the central region where the resolution is  lower
($r<2\epsilon_{g}$).

We first analyzed the density profiles of dark matter halos in purely
dynamical runs (I.1 and II.1).
We found that these halos
 fit the NFW profile
remarkably
well. So their density profiles  steep up as $r^{-1}$ in the central region
 as
predicted by NFW and Hern.
 We also observed that  more massive halos are less concentrated than
smaller ones. The concentration parameters $c_{NFW}$ found for our
halos are consistent with the values published by NFW (Table II). Fig. 4
shows four halo density profiles in a N-body run (I.1) and the
corresponding fitting using NFW model (solid and dashed lines respectively).

\begin{figure*}
\centering \psfig{file=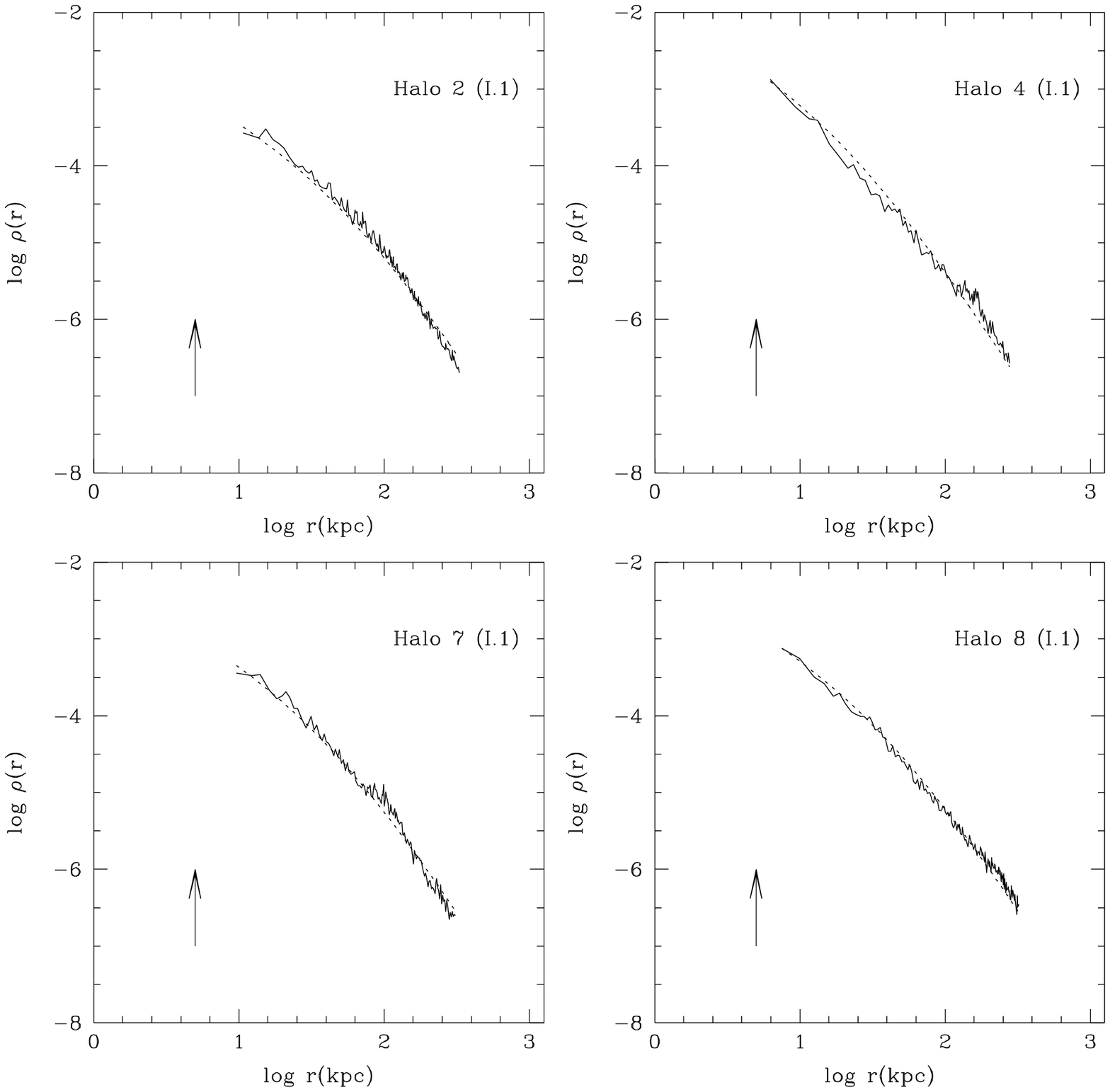,height=11cm,clips=,angle=0}
\caption{ }
\end{figure*}

Following the same procedure, we calculated the density profiles of
dark matter halos
in the hydrodynamical runs (simulations I.2 , I.3 and II.2).
These halos show some differences as can
be seen in Fig. 5, where the density profiles of halos 1 and 3
in simulations I.1 and I.2 have been plotted.
The solid line represents the density profiles in the hydrodynamical run
and the dashed lines the corresponding profiles in the purely dynamical one.
Since the simulations share the same initial conditions for both, velocities
and positions, and gravitational parameters (softening, force errors),
the differences found among the density profiles may
be directly related with  hydrodynamical effects.

\begin{figure*}
\centering \psfig{file=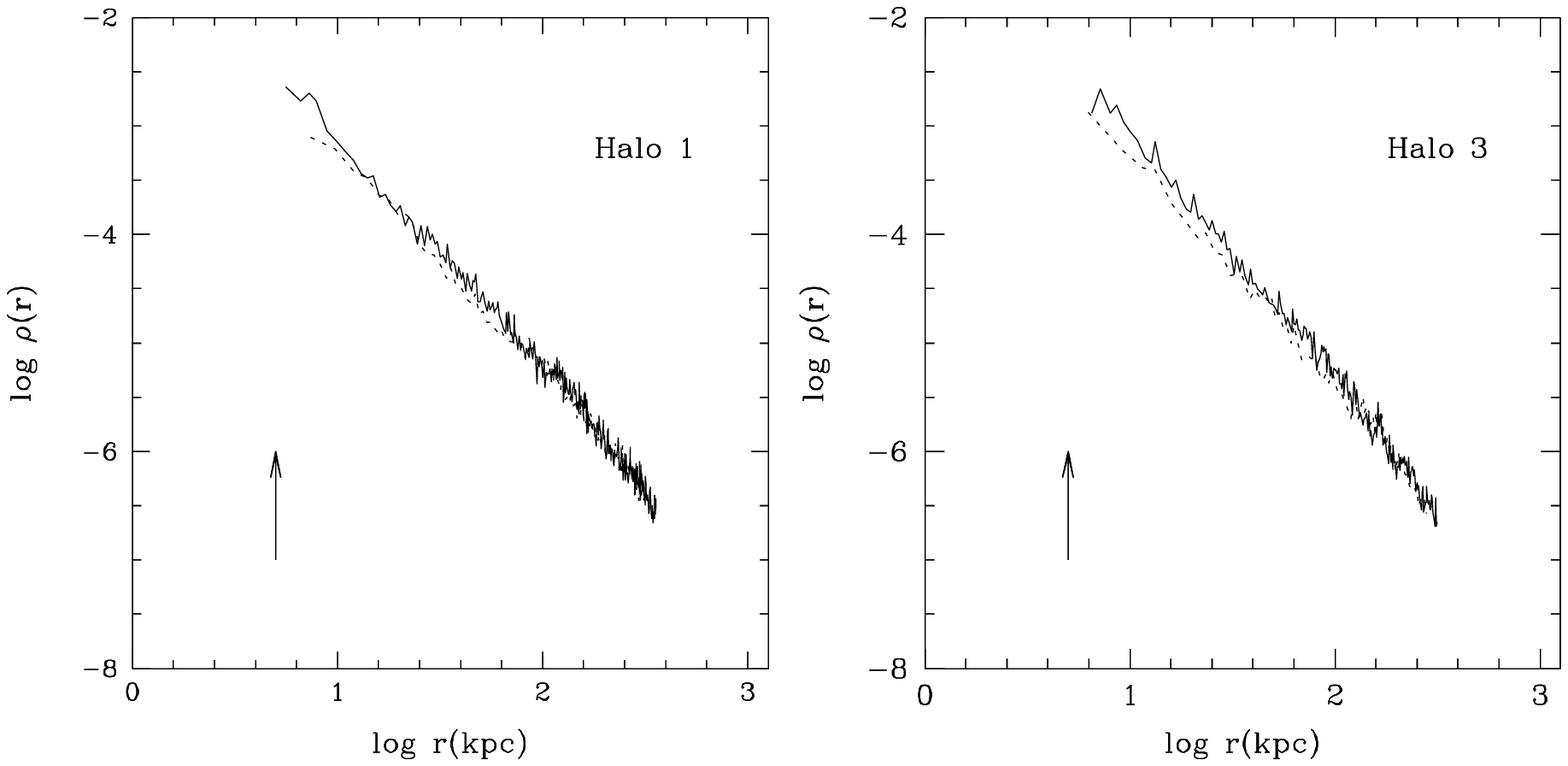,height=15cm,rheight=10cm,clips=,angle=0}
\caption{ }
\end{figure*}

These figures show clearly how the density changes because of the
presence of a baryonic core at the center of the halo.
In Fig. 6, we plot the density profiles of halos 1, 2, 3, 4 in simulation
I.2 (solid lines). We also include the best fit obtained using NFW profiles
(dot-long dashed lines).
The slope of the density  profiles of these halos
is steeper in the central region, so
the profiles defined by Navarro \eti (1995) are too curved  to fit them
 \footnote{The parameter $\delta_{c}$ has
been redefined for consistency in a similar way as in equation 6: $\delta_{c}=
\frac{c^{3}M^{dark}_{200}\rho_{crit}^{-1}}{3v_{200}[ln(1+c)-c/(1+c)]}$,
where $
M^{dark}_{200}$ is the dark matter within the virial radius and $v_{200}$
is the volume defined by the virial radius}.
Our  profiles seem to follow a simple power law  over a larger range of radius than
NFW profiles.
The radius at which NFW profiles depart from our simulated
profiles correlates with the radius of the galaxy-like object in
the halo (we define $r_{b}$   as the radius at which the baryonic
and dark mattter density profiles intersect each other. See section 3.3.1 for
details).
 It has to be stressed that the
dynamical resolution of our halos is comparable with NFW objects
for the same range of circular velocities.

\begin{figure*}
\centering \psfig{file=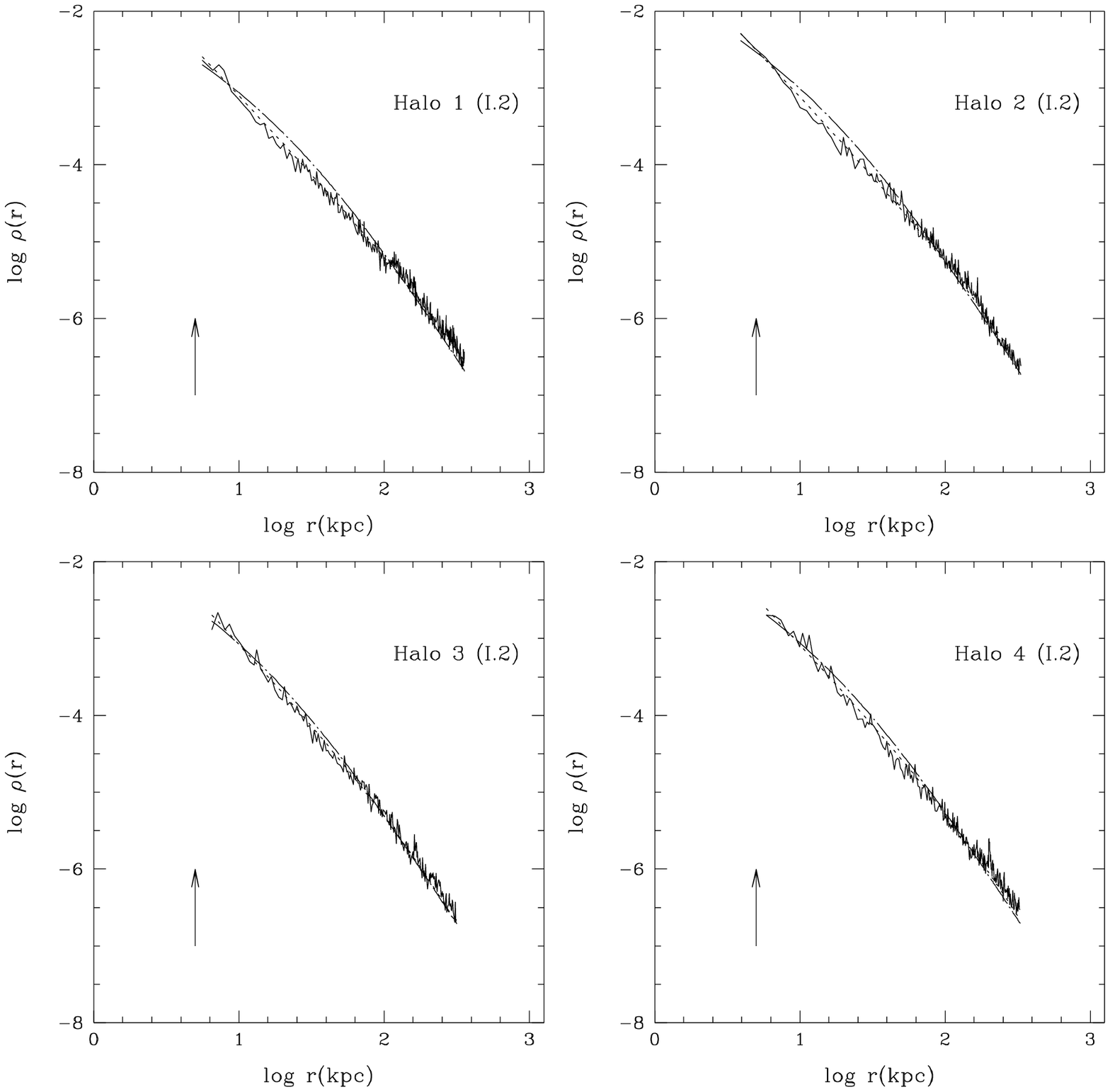,height=11cm,clips=,angle=0}
\caption{ }
\end{figure*}

 We use a standard algorithm to fit to these profiles a
power law in the central region ($\rho \propto r^{\alpha}$),  looking for the value of $\alpha$ that gives the best fit.
Values ranging from -1.9 to -2 where found. Hence, we decided to keep
the functional form defined by equation 1, but increasing the
slope of the power law for small radius:

\be
\rho(r)=\frac{{A_{c}}}{(r/r_{s})^{2}(r/r_{s}+1)^{2}}
\ee

where  $r_{s}=r_{200}/c_{TD}$ and the
$c_{TD}$ is the paramenter defined by NFW as the concentration
\footnote{ This is, in fact, the functional form studied by Jaffe (1983) for
spherical systems.}.
The parameter $A_{c}$ is determined by requiring that the total density
contrast
at the virial radius equals $200\rho_{crit}$. For equation (6) we obtain:

\be
{A_{c}}=\frac{\rho_{dark}}{3}c_{TD}^{2}(1+c_{TD})
\ee

where $\rho_{dark}= M^{200}_{dark}/v_{200}$. 
The mean dark matter density at $r_{200}$, $\rho_{dark}$, can be rewritten as 
$\rho_{dark}=(1-f_{b})
200\rho_{crit}$ where $f_{b}$ is the fraction of baryons within $r_{200}$.
This fraction is almost constant ($\approx \Omega_{b}$)
for all halos.
The profile defined by equation (6) was fitted to all halos in the hydrodynamical simulations.
In Fig. 6, we also plot the best fit for halos 1, 2, 3, and 4 using equation (6) (dotted
lines).
The parameters obtained are summarized in Table II.
As can be seen from this table, $c_{TD}$  values are smaller than
$c_{NFW}$ ones,  implying larger values of $r_{s}$ for halos in hydrodynamical runs.
 This can be understood
taking into account that, in equation (6), $r_{s}$ is the radius that sets
the transition from
$\rho(r)\propto r^{-4}$ for larger radius to $\rho (r) \propto r^{-2}$
for the intermediate region, while, in NFW,  $r_{s}$ delimites the change
in the slope from $\rho(r)\propto r^{-3}$ to $\rho(r)\propto r^{-1}$
in the central region.

\begin{figure*}
\centering \psfig{file=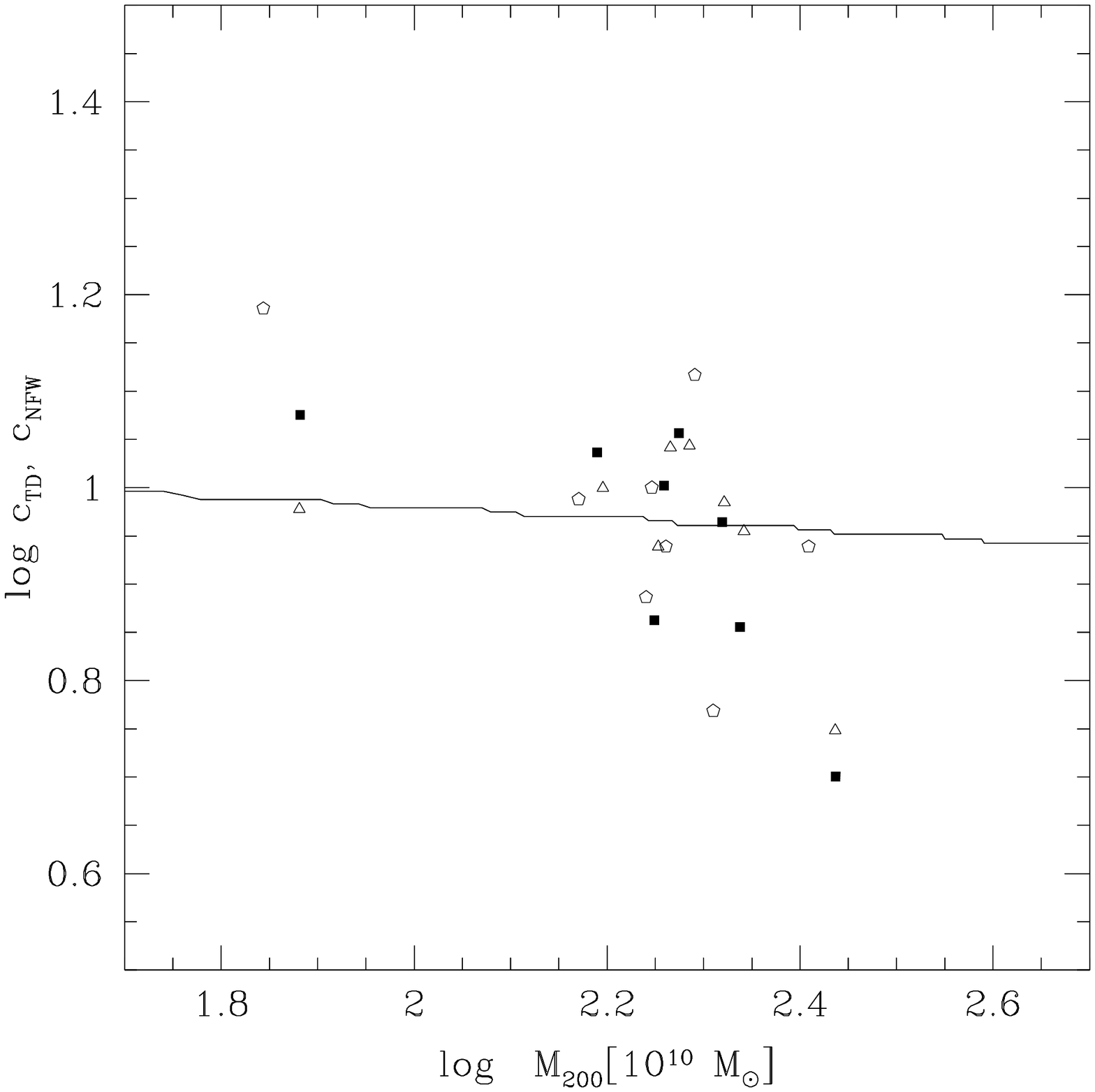,height=5cm,rheight=3cm,clips=,angle=0}
\caption{ }
\end{figure*}

Fig. 7 shows the correlation found between the logarithm of
  $c_{TD}$ and $M_{200}$
 for the density profiles of halos
in simulations I.1 (open pentagons), I.2 (solid squares) and I.3
(open triangles). The zero point of the relation for objects in I.2 and I.3
has been redefined  by adding an arbitrary constant to the logarithm of
$ c_{TD}$ in order to place them in
the same region of the diagram as the logarithm of $c_{NFW}$.
This figure shows that
larger masses are less concentrated, as pointed out by NFW
for purely dynamical simulations.
 We also include the theoretical predictions for $c_{NFW}$ from
the collapse time  assuming NFW density profiles (we used an algorithm kindly
provided by these authors to estimate the collapse time of the halos).
With this figure we intend
only to compare the slope of the relations for the hydrodynamical and N-body
runs. The  slope of the correlation for
 halos in I.2 and I.3 seems to be consistent with
the relation obtained from  the purely dynamical simulations.
 Note, however, that the range
of masses is small to get a general trend.
We did not find a difference between halos hosting a disk or a spheroid.
In any case, a large sample with higher hydrodynamical resolution and
dynamical range
is needed to have a more robust statistical signal.

\begin{figure*}
\centering \psfig{file=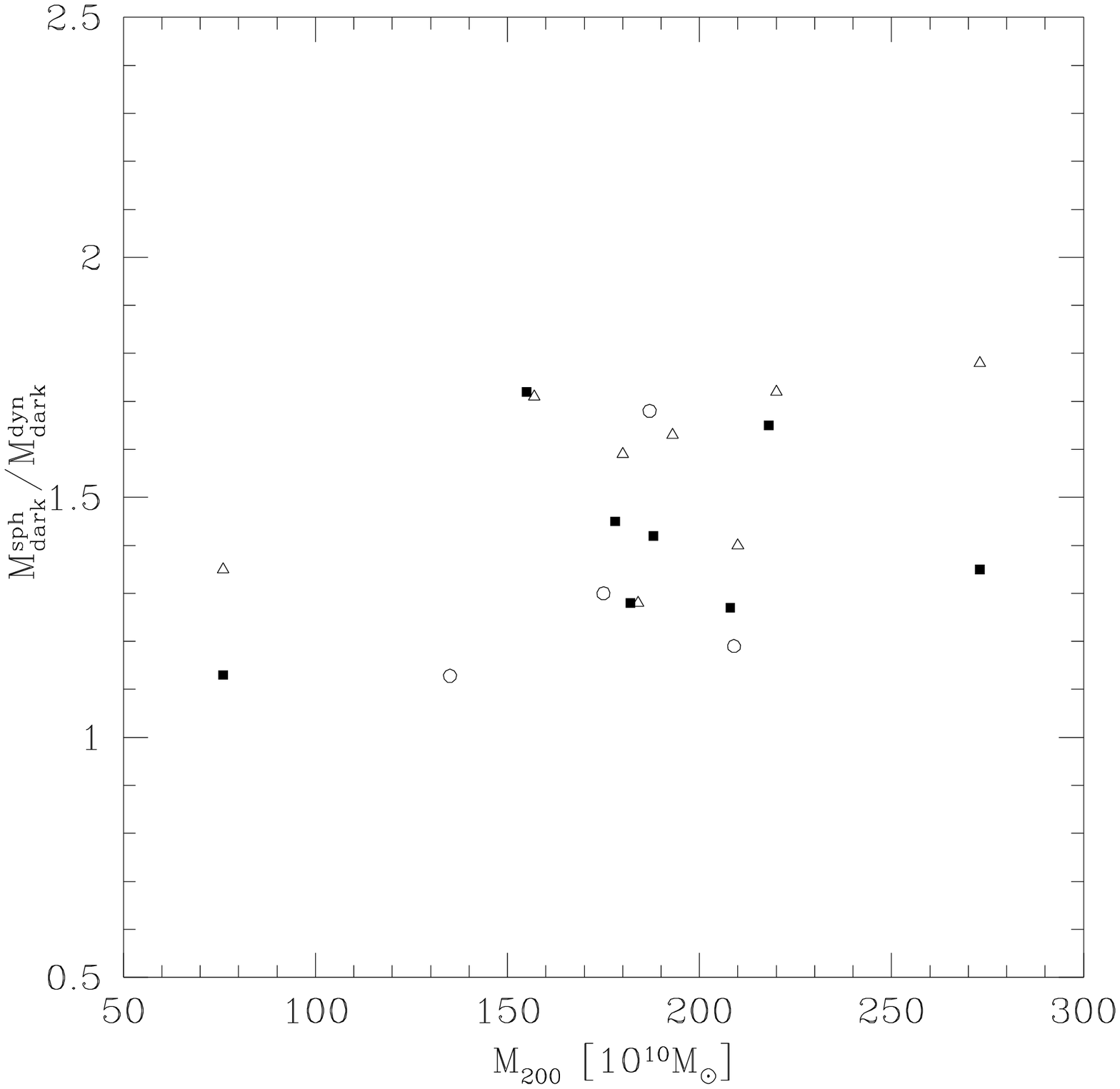,height=11cm,clips=,angle=0}
\caption{ }
\end{figure*}

We measured the ratio $\frac{M_{dark}^{sph}}{M_{dark}^{dyn}}$ where
$M_{dark}^{sph}$
and $M_{dark}^{dyn}$ are the dark matter masses for halos in the hydrodynamical
and the purely dynamical cases within a radius of $r=30$ kpc, respectively.
 This ratio roughly estimates the pulling of the dark matter by
the baryons (Blumenthal \eti 1986).
In Fig. 8 we plot it versus $M_{200}$. As can be seen from this
figure,  it does not seem to correlate
with mass.
The fact that it does not show a correlation with the mass
in the range of velocities we are
working with, indicates that all halos in our sample seem to have been
squeezed by approximately the same factor due to the presence of the baryons.
If it is valid, then  the slope of the correlation between the concentration
 parameters and $M_{200}$ could be still
determined by the collapse time of the object.

At this point, it is worth  noting that, in the absence of heating sources
(like supernovae energy injection) the baryonic matter
overcools at high redshifts leading to
the formation of highly concentrated objects (see for example
Navarro \& Steinmetz  1996). This
effect may also affect the evolution of the dark matter in
the central region of the halos by increasing artificially
the concentration of
particles. In spite of this fact, as discussed by other authors, the
presence of baryons does change the matter distribution and this is
clearly observed in our simulations, although the effect may be not as strong
as measured here. We will come back to this point in the next sections.

\subsubsection{Baryonic Density Profiles}

We construct the  density profiles for the baryonic matter belonging to each 
halo listed in Table II by following the same procedure
described in section 3.3 for the dark matter component.
The density profiles of baryons are not as accurate as the dark matter ones
since
the number of particles used to solve the gaseous
component is  lower than the one used for the dark matter.

As an illustration of the general behaviour of the baryonic component,
Fig. 9 shows the density profiles for baryons and dark matter for two halos
in set I. These two profiles intersect
each other at a certain radius, $r_{b}$, which can be taken as 
a measure of the physical size of 
the central galaxy-like object. 
 The same is true for the other GLOs listed in Table II.
The $r_{b}$ values range from 14  kpc to 30 kpc,
but they are difficult to be properly determined because of the noisy
character of the baryonic density profiles.

\begin{figure*}
\centering \psfig{file=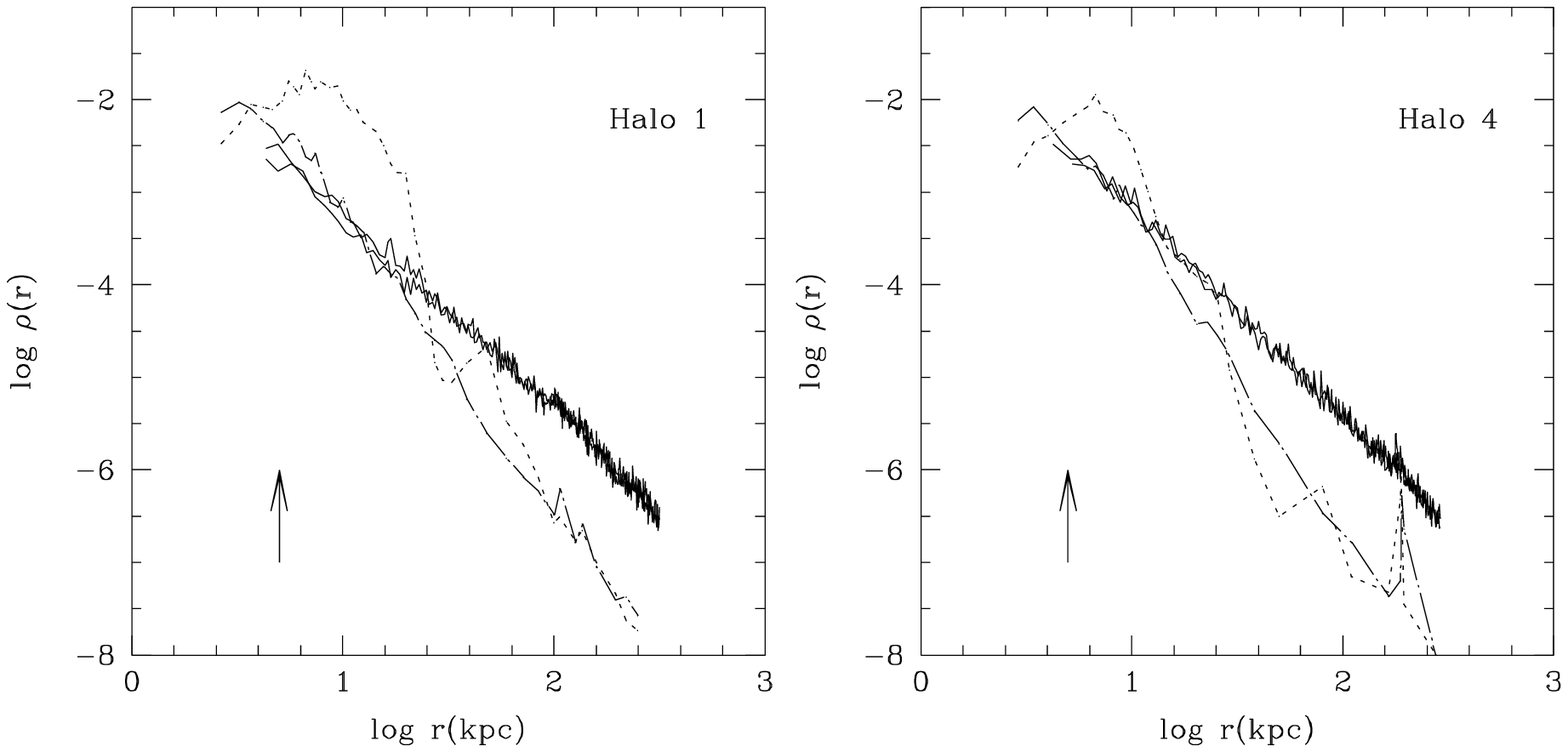,height=11cm,rheight=6cm,clips=,angle=0}
\caption{ }
\end{figure*}

 Note  first that the baryonic density profiles (dot-long dashed and
dotted lines) are steeper than
the dark matter ones (solid lines).
The slope for radius greater than $r_{b}$ is
$\rho_{bar}(r) \propto r^{-3}$. Secondly, that the baryonic dark matter profiles
in the central region  depend on the star formation parameters used.
In simulation I.2 (dotted lines), baryons seem to have collapsed more dramatically
than in I.3 . In this last run (dot-long dashed lines), the baryonic density profiles
 follow the dark matter for $r< r_{b}$, changing  its slope to
$\rho_{bar}(r) \propto r^{-2}$. This is consistent with the fact that the gas has been more
efficiently transformed into stars and that stars behave in a similar
fashion as dark matter particles since they are only affected by
gravitational forces.
On the other hand, in I.2 the higher concentration of baryons in the
central region is the consequence of a higher efficiency in the cooling
of the gas particles. Hence, the history of star formation
is relevant to the final
distribution of baryons in the central region.
However, from our simulations we cannot detect a clear change in
the dark matter density profiles (solid lines) as a result of the difference in the
baryonic distributions.


\subsubsection{Circular Velocities}

The first consequence of the difference between dark matter density profiles of halos
in hydrodynamical simulations and purely gravitational ones is that
 the maximum of the circular velocity curves  ($V_{cir}^{2}(r) = GM(r)/r$) 
 of these
halos translates towards smaller radii.
In simulation I.1, $r_{dark}^{max}$ is
around $50-80$ kpc.
Whilst in simulations I.2 and I.3, $r_{dark}^{max}$ is found at approximately
$20-30$  kpc  .

\begin{figure*}
\centering \psfig{file=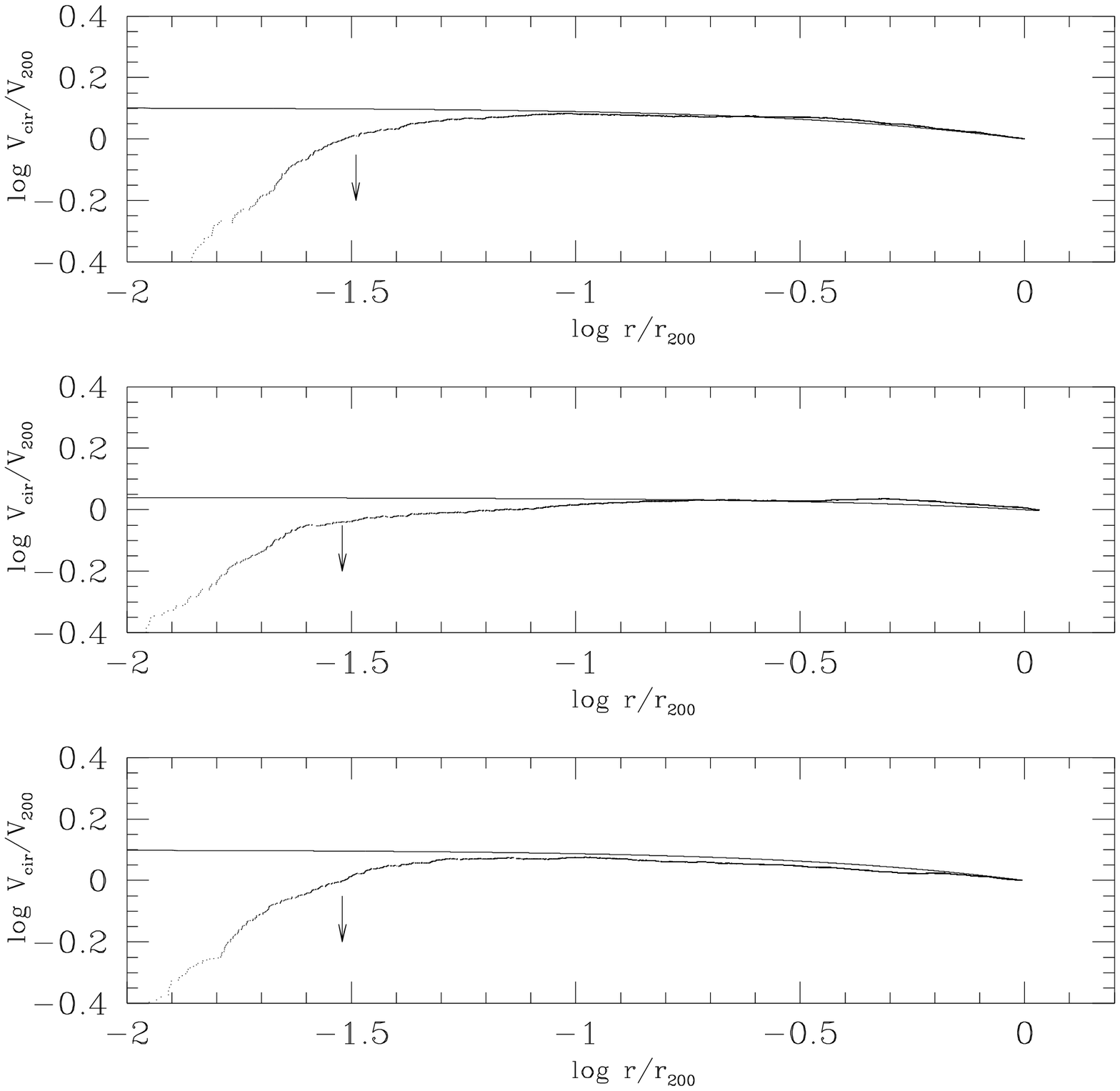,height=11cm,clips=,angle=0}
\caption{ }
\end{figure*}

According to equation (6), the circular velocity curve of dark matter halos
is given by :
\be
V_{cir}(r) = \frac{V_{200}(1+c_{TD})^{1/2}}{(1+\frac{r}{r_{s}})^{1/2}}
\ee

where $V_{200}$ is the circular velocity of the dark matter
at the virial radius.
This curve agrees with the circular velocity for an isothermal
sphere for inner radii, but decreases as $r^{-0.5}$ for larger ones.
As an example, Fig. 10 shows $V_{cir}(r)$ for 
halos 1, 2 and 4 in simulation I.2 and the fitting
using equation (8). As can be seen from this figure, the analytical fit
is very good except for small radius where the effects of numerical
resolution on the integrated mass is noticeable
(the arrows point out $r=2\epsilon_{g}$).
Our numerical resolution does not allow us to properly described
the density profiles of either, the dark matter  or the baryons, within
$5$ kpc.
Numerical simulations with much higher resolution would be needed in
 order to  assess the slope of distribution of dark matter
in the very central region.

\subsection{Rotational Properties}
The problem of understanding the rotation velocity curves
of spiral galaxies has been analyzed in several works from
 observational (Persic \eti 1996 (PSS) and references therein)
and theoretical points of view
(Blumenthal \eti 1986, Flores \eti 1993). In this paper,
we study the rotational properties of disk-like objects formed in
 cosmological hydrodynamical simulations and  whose
halos were previously analyzed in section 3.3.

\begin{figure*}
\centering \psfig{file=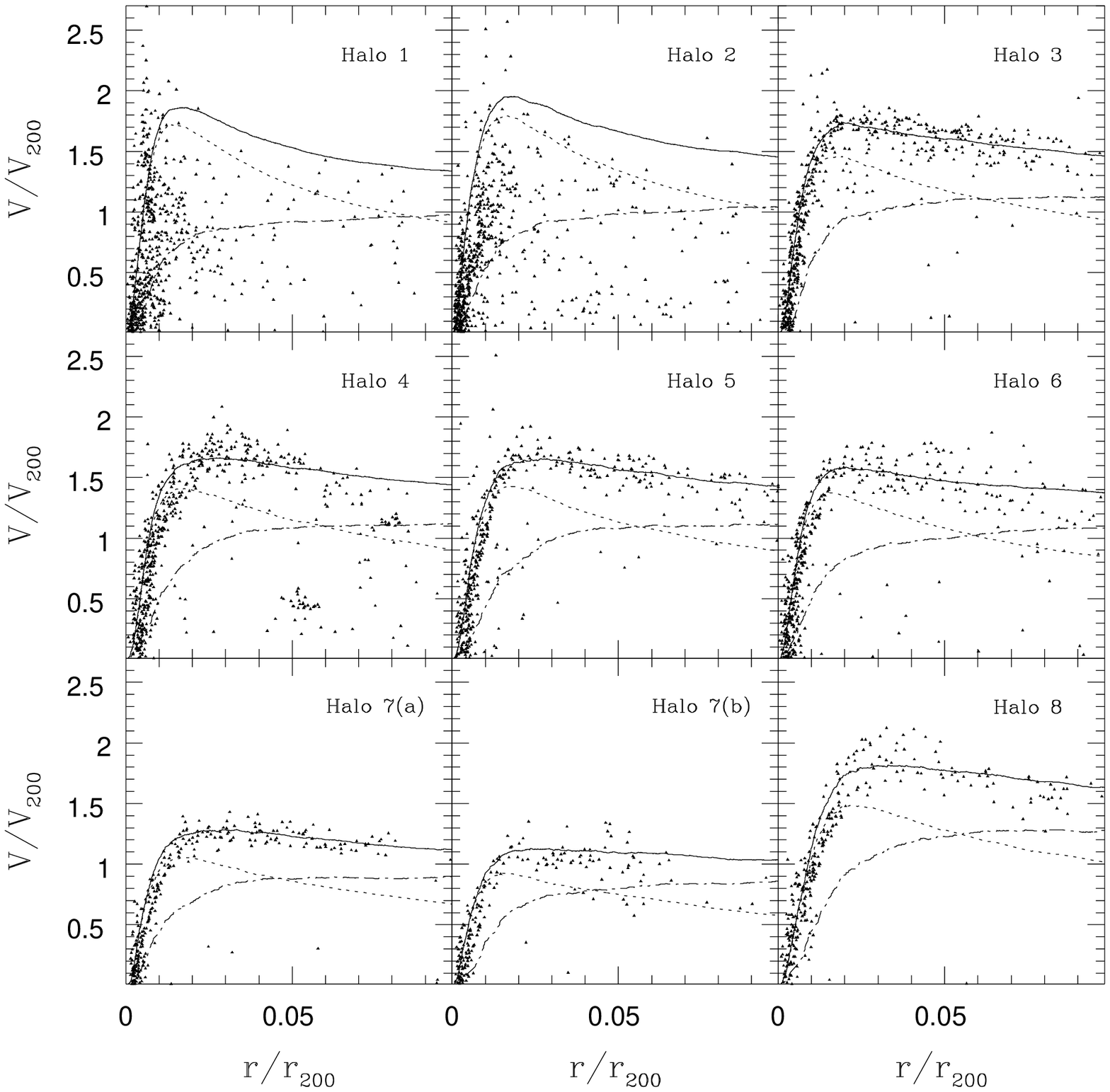,height=11cm,clips=,angle=0}
\caption{ }
\end{figure*}

We construct the rotation velocity distributions of the
disk-like structures formed in  halos in simulation I.2 by calculating
the tangential velocity component of the baryonic particles belonging to the disks,
in a cylindric reference system defined by the
direction of the angular momentum of the disk with origin
at its center of mass. Fig. 11 shows the rotation velocity distributions
  for  7
disk-like objects   (solid triangles; recall that halo 7 hosts a pair of disks).
 Two extra objects have been included which are not
purely disk-like structures but spheroids and are more massive than
any of the disks shown (halos  1 and 2).
We also include in this figure the
 equilibrium circular values of
the velocities for the total mass (solid lines).
We adopted
the softened Plummer potential (Evrard \eti 1994):
\be
V_{cir}^{2}(r)= \frac{G M(r)r^{2}}{(r^{2}+\epsilon _{g}^{2})^{3/2}}
\ee

As can be seen from this figure, baryonic particles in disk-like objects
are in coherent rotation with tangential velocities which show small dispersions
about the $V_{cir}(r)$ curves,
reflecting the fact that they  are in rotational supported equilibrium within
the potential well of the dark halos.
For the spheroids, the tangential velocities appear to be
disordered, showing that no disk coherent rotation exits.
We also plotted the circular
velocities of the dark (dot-dashed lines) and  baryonic (dotted lines)
components separetely.

In simulation I.3 no disks have been formed, so
we look at the circular velocity of baryons and dark matter of
the objects  as an indicator of mass distribution
and for comparison, in order to assess differences mainly due to a
different  star formation history (Fig. 12). We find
interesting differences when equivalent galaxy-like objects
in simulations I.2 and I.3
are compared.  First of all, as already mentioned,
 the percentage of remanent gas
 in run I.3 is much lower and there are  no clear disk-like structures.
 The gas distributions are spheroidal structures
 with $<c/a> \approx 0.5$.
 The two baryonic objects which are spheroids in simulation I.2
   are also spheroids in run I.3.
   The rest of the objects has
a tenuous gas disk. In any case, the baryon fraction at a given $r/r_{200}$ 
in the inner regions 
is lower for GLOs in simulation I.3, as can be seen from Figs. 11 and 12.

\subsection{Baryonic and Dark Matter Distributions in the Inner Regions}

An interesting observation coming out from Figs. 11 and 12 is that the
baryonic and dark matter distributions cross each other at a certain
radius (crossing radius, $r_{cross}$). This radius is the limit which
divides the baryonic dominated region from the dark matter one.
Hence, the rotation curves are  dominated by
the baryonic matter for $r< r_{cross}$, while the dark component is
preponderant for larger radius. The values of $r_{cross}$ for GLOs in
hydrodynamical runs are given in Table II. Recall that the GLOs shown 
in Figs. 11 and 12 are exactly the same, but with different 
star formation histories (Table I)

The first question to be addressed is whether the halos and the GLOs
they host present a self-similar mass distribution, that is,
whether the baryon density profiles are self-similar when normalized
to the parameters $r_{s}$ and $A_{c}$ that determine the
self-similar character of the dark matter profiles (see section 3.3).
A positive answer would mean that the baryon distribution is mainly
determined, at any scale, by the dynamics of the dark matter halos, and
would imply that $r_{cross}/r_{s}$ is the same for all the GLOs
belonging to a given run. This is not the case, as can be deduced
from Table II, 
suggesting that, at small scales, the  baryon distributions are
linked to some other physical processes.

In order to study these distributions, we have analyzed the
possible correlations of $r_{cross}/r_{200}$
(the fraction of the virial volume dominated by baryons) with 
a global halo dynamical parameter ($M_{200}$ or $ V_{200}$),
 on one hand, and with  local
measures of the baryon distribution at the central region of the GLOs, 
on the other.
Crossing radii normalized to the virial radii have been used instead of the
crossing radii themselves to properly assess any effect that could arise
with independence on the halo size.

\begin{figure*}
\centering \psfig{file=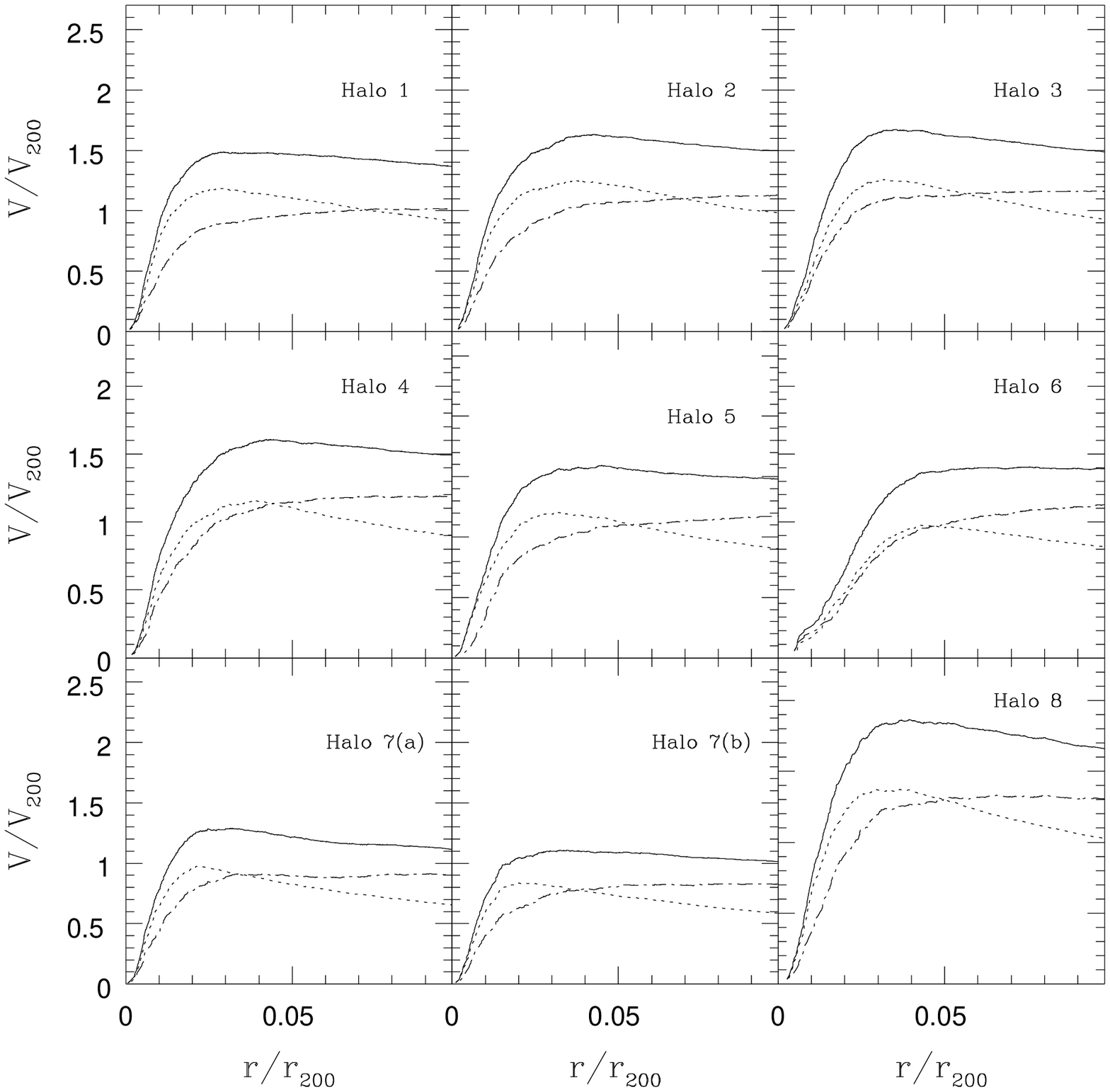,height=11cm,clips=,angle=0}
\caption{ }
\end{figure*}

In Fig. 13a we plot  $V_{200}$ versus 
$r_{cross}/r_{200}$, finding
 no clear correlation between these parameters. 
However, halos with similar virial velocities are found to have 
different $r_{cross}/r_{200}$. This fact implies that their 
 inner regions are dominated by baryons at a different extent.

In Fig. 13b we plot  
the maximum of the circular velocity curve, $V_{cir}^{max}$ (see equation 9),
versus $r_{cross}/r_{200}$ for
GLOs shown in Figs. 11 and 12. This graphic shows a correlation for objects
in simulations I.2 and I.3.
The meaning of this correlation  is the following. On one hand, $V_{cir}^{max}$
is a measure of the total mass within a sphere of radius $r_{bar}^{max}$ 
(that,
as seen from Figs. 11 and 12, is mainly made of baryons) and of
the compactness of its distribution.
On the other hand,
$r_{cross}/ r_{200}$ increases when the fraction of baryons to
dark matter in the inner regions
of the rescaled version of the GLOs increases. So the correlation merely 
expresses the fact that an increase in the baryon fraction in the
inner regions of the GLOs (normalized to their total size) is linked to
an increase in the total (baryon plus dark) mass inside  $r_{bar}^{max}$,
and/or the compactness of its distribution.
A more detailed analysis of this very inner region of the GLOs is not possible
to carry out since they lie within the dynamical resolution of the simulations.\

Note from Figs. 11 and 12 and Table II that $r_{cross}/r_{200}$  is
smaller for GLOs in simulation I.3 than for their counterparts in run I.2,
and even smaller for objects in run II.2.
In Fig. 13(a,b) these differences are also noticeable. 
We think that these differences in the values of the crossing
radii  are due to the fact that
the gas has been transformed into stars more efficiently and from
earlier times in I.3, and, when star particles are formed, they are instantaneoulsy
treated as collisionless. Hence, the collapse of the cold gas towards the
centre of the object might no have been that effective in this simulation.
This same effect can also explain the small 
$r_{cross}/r_{200}$
 values in GLOs belonging to run II.2, where the star formation process
has started even at higher redshifts.

A related question is whether or not   the fraction
of the total baryonic mass inside the virial radii of the halos that is
 accreted at  their centre  depends on the virial mass of the halos.
 To clarify this point, we have also 
plotted the ratio $M_{bar}^{cross}/ M_{bar}^{200}$ versus $M_{200}$
 (Fig. 14). As can be seen from this
figure,  no correlation has been found
for objects belonging to the same run, or, at least, it seems that 
it would be too weak to
be detected with our sample.
This figure  shows that the fraction of baryons in the central regions of
the GLOs grows from run II.2 to I.2. This effect
is equivalent to the differences
found for $r_{cross}/r_{200}$ among objects belonging to different runs,
and can be explained taking into
account the differences in their star formation histories as discussed above.

We now try to understand how the results  just described are originated. 
Let us first consider how numerical resolution can affect the infall
of baryonic matter.
According to Navarro and Steinmetz (1996), gaseous disk-like objects resolved
with a few hundred of particles have their gas density artificially
smoothed and the accretion and collapse of the disk is delayed.
However, the effects measured by these authors could be larger than
the ones present in our simulations since, in their low resolution
experiments, the authors diminshed the resolution of both, the 
dark matter and baryons. Taking into account  the results of
Steinmetz and White (1997), this is  the scenario that produces the
larger numerical artifacts, since also the potential wells of the halos
are poorly represented. In our simulations, the potential wells onto where the
baryons collapse are   described with similar resolution to 
the high resolution runs of Navarro and Steinmetz (1996).
On the other hand, the objects chosen  from our simulations to be analyzed are resolved
with several hundred of gas particles, varying by a factor of 2 (Table II).
So, larger objects are slightly better resolved than  smaller ones.
Nevertheless, because of the similar number of gas
particles used, we would tend to think that with our hydrodynamical resolution
limits, all objects are likely to be affected at the same level. 
Note  also that Fig. 14 shows no correlation for this sample, implying
no dependence of the baryonic infall within the crossing radius 
on the halo virial  mass.
Secondly, we consider the possible effects of overcooling. Taking into account 
the relation between the cooling time and the dynamical time
of a collapsing object, the cooling process are  more effective
at higher redshifts, producing the early collapse of baryons onto
the potential wells of the first halos. Hence, overcooling effects
would  lead to a correlation we do not find,
 since smaller baryonic objects would
be more affected than larger ones which have a later collapse, and, 
consequently, would have  a higher fraction of baryons at their centres.

\begin{figure*}

\centering \psfig{file=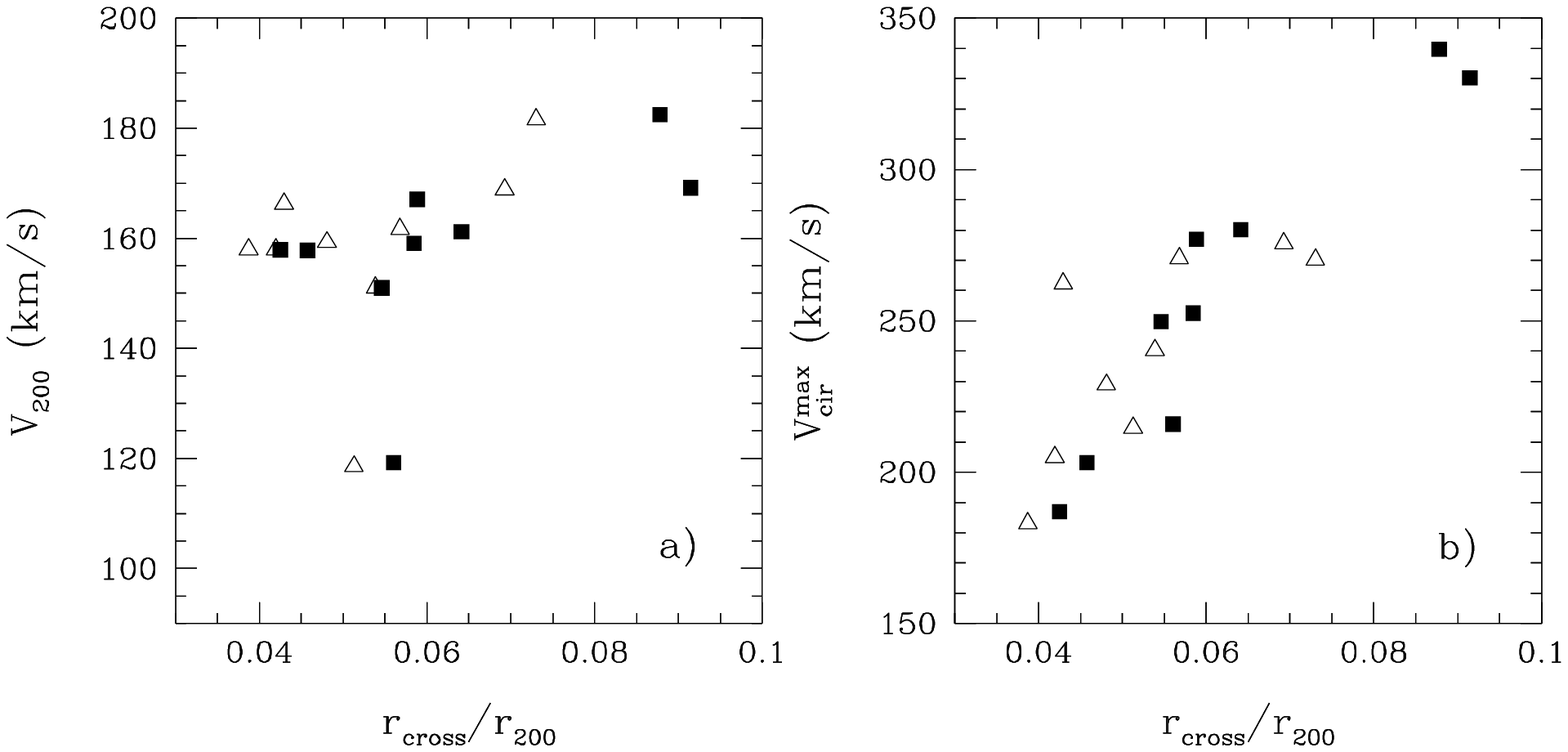,height=13cm,rheight=7cm,clips=,angle=0}
\caption{ }
\end{figure*}

\begin{figure*}
\centering \psfig{file=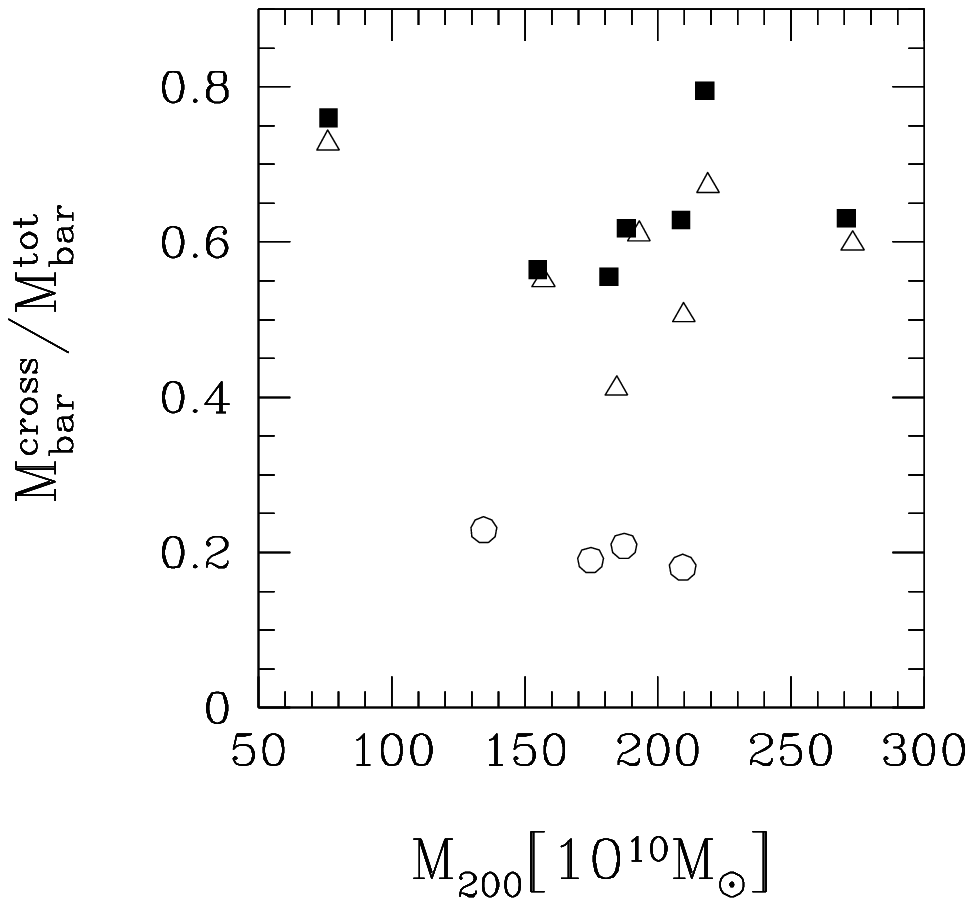,height=13cm,rheight=7cm,clips=,angle=0}
\caption{ }
\end{figure*}

So, we need to invoque different mechanisms
 to explain Figs. 13(a,b) and 14.
Let us now consider the  characteristic evolutionary path  a GLO 
goes over in a
hierachical clustering scenario, where they are
built up through  mergers of smaller baryonic units composed of gas and stars.
These mergers take place whithin dark matter halos, that have previously
suffered their own mergers and virialized, and are characterized by
parameters such as the structure, mass, relative orientation, 
orbital angular momentum and orbital energy
of the two baryonic clumps that 
are to merge (Navarro \eti 1995). These parameters vary in each 
case, and for each merger process, they vary  along the time 
in a complicated manner.
 So, each GLO has its own merger tree of
baryonic clumps, that, in addition, may determine in part its star formation
history (Tissera \eti 1996). These trees may be (and in most cases they are)
different for different GLOs, even if the global
dynamical characteristics of the halos that host them are similar.
More precisely,
some authors (Hernquist 1989,
 Barnes \& Hernquist 1991, 1996)
have suggested that during a merger process of baryonic units, a
gas inflow is expected to be drawn towards the central region as a consequence
of a transference of angular momentum from the gas component to the 
collisionless matter.
In particular, Barnes \& Hernquist (1991) studied the dynamics of gas
in a merger using a N-body/hydrodynamical code without star formation.
They found that the gas component in the inner half of each disk loses most
of its angular momentum and falls into a compact cloud at the center
of the galaxy. 
Even if our simulations are not directly comparable 
to these results, mainly because star formation results in bulge-like
baryonic cores that could slow down the gas inflow rates (Mihos \&
Hernquist 1994), they suggest that mergers of baryonic clumps
can be highly efficient in driving gas towards the center of galactic objects.
This suggests the possible existence of a  correlation between the number
and the characteristics 
of mergers suffered by a baryonic object and its crossing radius, and,
moreover, the absence of correlation between this quantity and
the virial mass of the halos. 
 So mergers might be considered
as an effective mechanism to redistribute the gas component
within the radius of a galactic object leading to more concentrated objects.
Hence, the combined history of star formation and mergers seems to be relevant
to determine the distribution of baryons in the inner regions of halos and, 
therefore, the properties of the rotation curves of GLOs.   
Phenomenologically, this translates into the need for additional parameter(s),
other than a halo scale and a halo characteristic density, 
to describe these rotation 
curves, in accordance with the  findings of PSS. These authors claim that
an extra parameter (the galaxy luminosity) is needed to 
characterize the shape of
the {\it normalized} rotation curves of the galaxy sample they have studied.

\subsubsection{Comparison to Observations}

PSS point out that spiral
galaxies in their sample with $V_{rot} > 200 \ \rm km \ s^{-1}$ have decreasing 
rotation curves, while
less massive galaxies have almost flat rotation curves. 
These authors derived from  observations of rotation velocities and
luminosities that the ratio between dark matter and baryonic matter
inside the optical radius (defined as the radius that encloses $83 \%$
of the luminous mass of the disk),
was an inversely increasing function of
velocity. Thus, the inner part of  high luminosity spirals seems to be
 dominated
by luminous matter, while, conversely, low luminosity objects are dominated
by dark matter.
Bigger disks in our simulations (in the sense that they are more massive at
radii smaller than $r_{cross}$)
tend to have larger crossing radii, in agreement with
the results of PSS.
Hence, the kinematics of  massive disks seems to be strongly determined
by the luminous mass. 

\begin{figure*}
\centering \psfig{file=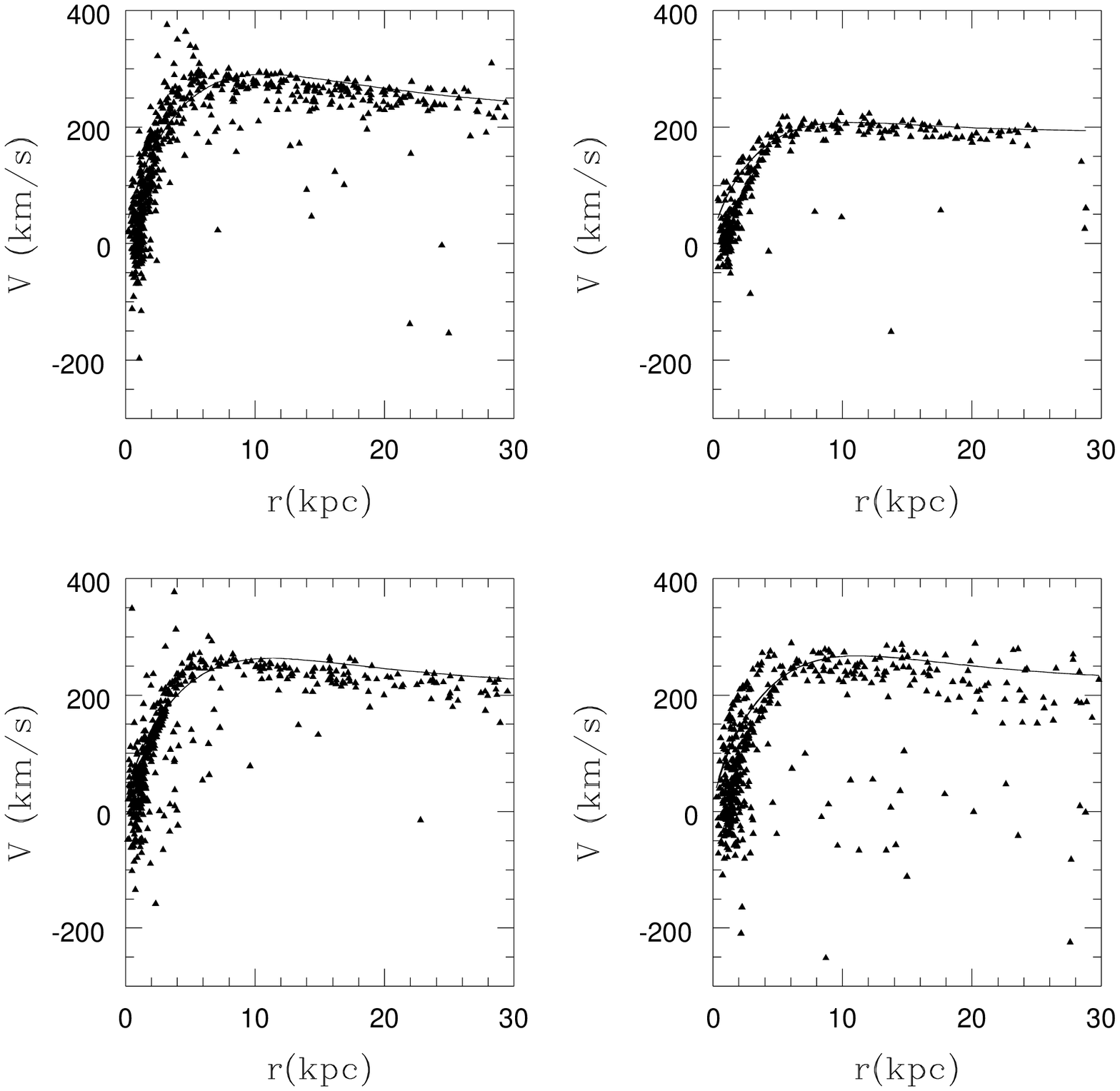,height=11cm,clips=,angle=0}
\caption{ }
\end{figure*}

We attemp to fit the circular velocity curve of
the  disks shown in Fig.  11 using the functional relation
found by PSS (Universal Rotation Curve, URC). 
Their observational
 sample is composed
by late-type spirals with almost no bulge contribution. So, the baryonic
mass distribution is modelled as an exponential disk.
These authors assumed a simple density profile for the dark matter with a core.
The resulting rotation curve is obtained as a sum of the disk and
halo contribution  whose parameters are adjusted using  observational data.
They obtain:

\begin{eqnarray}
V_{URC}(R/R_{opt})
&=V(R_{opt})[ (0.72+0.44{\rm log}\frac{L}{L_{*}})
\frac{1.97x^{1.22}}{(x^{2}+0.78^{2})^{1.43}}\cr \nonumber
& \qquad  +1.6{\rm exp}[-0.4(L/L_{*})]\frac{x^{2}}{x^{2}+1.5^{2}
(\frac{L}{L_{*}})^{0.4}}]^{1/2}
\end{eqnarray}

where $x=R/R_{opt}$, and
$R_{opt}$, $V_{opt}$ and $L$
are the optical radius, the velocity at the optical radius
 and the  luminosity in the B-band.

Eventhough the hypotheses assumed by these
authors do not agree with the properties of the objects identified in our
simulations, namely, they  have a bulge contribution and the
dark matter density profiles are singular in the centre, we have been able
to fit the URC proposed by PSS to the circular velocity curves of our disks
(see Fig. 15). The {\it shape} of the simulated disk-like objects
is not in disagreement with the {\it shape} of the URC 
as shown in this figure for four disks. However, for
a simulated disk, the $L$ parameter can be also obtained from the
ratio of the baryonic to the total mass (baryons plus dark matter)
contributions to the circular velocities at $R_{opt}$ (see PSS for
details). The values of this parameter obtained by these
two methods are inconsistent. 
That is, the shapes of the rotation curves can be fitted
for a given set of the free parameters, but the 
values of these parameters 
are meaningless, implying that this agreement  does not
result from  underlying physical processes. This is not surprising,
given the incompatibility of the hypotheses in PSS and the characteristics
of the simulated disks we have obtained.
A detailed discussion of this issue
is delayed to a forthcoming paper.

\section{Summary}
We have carried out a comparative study of galactic objects in N-body
and hydrodynamical simulations paying special attention
to the analysis of the properties of the dark matter in galactic halos
in both kinds of scenarios.
Our main aim was to assess any differences in the evolution of the
dark matter caused  by the dissipative characteristic of
baryons. This kind of analysis presents several difficulties
  mainly
due to  numerical resolution limitations and the simplicity of the model used.
 Nevertheless, we think the results obtained clearly show the importance
of studying the joint evolution of both components in galactic-scale systems.

We arrive at the following results:

1- In agreement with other authors we find that density profiles of dark matter
halos in N-body simulations can be fitted by the analytical profile suggested
by NFW or Hern.

2- The consistent treatment of the joint evolution of dark matter and baryons
allows to detect a difference in the evolution of the dark component when
compared to purely dynamical models:

a- The shapes of dark matter halos change from prolate to more oblate
structures  when baryons are included.
 The dark matter halos in hydrodynamical simulations are
more prolate in inner region becoming more spherical at larger radius.
 A slight difference in shapes was measured when comparing halos which host a disk or a spheroidal baryonic
object. Nervertheless, a statistical analysis over a larger sample should be carried
out to confirm these results.

b- The anisotropy parameter $\beta_{dark}$ shows that dark matter halos are
nearly isotropic in the central region becoming more dominated by
radial dispersions at larger radius.

c- The total velocity dispersion of dark matter component is affected by
the presence of baryons. A stronger   increase of $\sigma$ is measured
 in the central region of all halos in hydrodynamical simulations than
in their purely gravitational counterparts.
We did not detect any dependence of $\sigma$ on the shape of the main
baryonic clump at the center of the halo.

d- The density profiles of dark matter halos in SPH simulations
are   fitted by neither
NFW nor Hern profiles. A steeper power law is required
  in the central region.
Instead
we fit all density profiles of dark matter halos in SPH simulations
with the curve defined by equation 1, but with the  combination of power:
$(\alpha,\beta,\gamma)= (2,1,2)$.
Hence, the profiles agree with the behavior of an isothermal sphere for
a larger interval of radius.
The ratio $\frac{M_{dark}^{sph}}{M_{dark}^{dyn}}$ indicates that approximatly
all our halos have been squeezed by the same factor, and that this
factor does not depend on their mass.

3- We estimate the density profiles of baryons finding a segregate behaviour
with respect to the dark matter. Baryons are more centrally concentrated.
We also observe that the slope
of the baryonic density profiles depends on the star formation parameters.
Those halos formed in simulation I.2 where a delayed star formation was
forced, are more  baryonic concentrated; whilst in the rest of the
halos, the baryonic density profiles followed the behavior
 of dark matter density profiles in the central region.
However, we found no differences among their dark matter density profiles.

4- We analyze the circular
velocity curves of the dark matter, baryons and the rotation
velocity curves for disk-like objects identified in SPH simulations.
 We find no
correlation between  the $r_{cross}/r_{200}$  and  the virial mass
of the halos. 
On the other hand, no systematic differences are found in the infall of baryons
within  $r_{cross}$ among the halos analyzed. However, we found a
clear correlation between the maximun of the circular  velocity curves and
$r_{cross}/r_{200}$. Hence,
this correlation seems to be originated in a different distribution
of baryons within the central regions. One possible mechanism which
could produce this effect are mergers with substructure.
The values of  $r_{cross}/r_{200}$  seem also to depend on the
history of star formation. If a large fraction of gas is transformed into
stars at early times, the objects are less dominated by baryons in the central
region since they cannot collapse so efficiently.

5- We have been able  to fit the URC  proposed
by PSS to the simulated rotation velocity
curves. We find a set of URC parameters which allows to match to the {\it shapes} of
the curves.  However, the luminosities of  simulated disks can also be
 found by considering the contributions
of baryons and dark matter to these curves. The two methods produce inconsistent
results, mainly due to the high concentration of baryons in the inner
regions of GLOs. So, the fits do not carry any physical information.

To sum up, the joint gravitational evolution of dark matter and baryons
in a hierarchical clustering model, together with physical processes
such as radiative cooling, star formation, supernovae, etc, may  be
all non-negligible interrelated ingredients to take into account when trying
to fully understand how disk galaxies formed, and in particular, to
explain their rotation curves.
In this paper, we analyzed the effects of introducing hydrodynamics and
star formation on the properties of galactic-like structures
 in CDM simulations.
There are several aspects such as the
star formation mechanism and processes related to it, dynamical resolution
and range,  which  would require to be improved in
future works.
Futhermore, overcooling effects are unavoidable without
 including heating sources
like supernova energy ejection. The properties of the
central region of all our halos
in the SPH runs may be affected by them.
Therefore, the models studied here and their outcomes are first
results from where underpin more sofisticated ones.
These facts will be discussed in future works where we will analyze
more complex models.

\section*{Acknowledgements}

We thank the referee of this paper for a careful reading and useful
comments and suggestions.
P. Tissera was supported by the Ministerio de Educaci\'on y Ciencia (Spain)
through a fellowship for foreign scientists.
P. Tissera would like to thank  Prof. Rowan-Robinson and the Astrophysics group at ICSTM for their
hospitality and support,  and to Prof. Efstathiou and
the Astrophysics group at the University of Oxford for
 providing the computational support
required to perform this paper.

\section*{References}

\noindent Aninos, P. , Norman, M. L., 1996, ApJ, 459, 12

\noindent Barnes J., Hernquist, 1996, ApJ, 471, 115

\noindent Barnes J., Hernquist, 1991, ApJ, 370, L65

\noindent Bertschinger, E., 1985, ApJS, 58, 39

\noindent Blumenthal G. R., Faber S. M., Flores R. A., Primack J. P., 1986 ApJ, 301, 27

\noindent Carlberg R. G., Lake G., Norman C. A., 1986, ApJ, 300, L1

\noindent Chieze, J., Teyssier, R., Alimi, J.-M, 1997, ApJ. Submitted

\noindent Cole S., Lacey C. G., 1996, MNRAS, 281, 716

\noindent Curir A., Diafero A., De Felice F., 1993, ApJ, 412, 70

\noindent Dubinski J., Carlberg R., 1991, ApJ, 378, 496

\noindent Dubinski, J., 1994, ApJ, 431, 617

\noindent Efstathiou, G. P., Frenk, C. S., White S.D.M., Davis M., 1988, MNRAS, 235, 715

\noindent Evrard, A.E., Summers, F. J., Davis, M. 1994, ApJ, 422, 11

\noindent Fillmore, J.A., Goldreich, P., 1984, ApJ, 281,1

\noindent Flores R. A., Primack J. R., 1994, ApJ, 427, L1

\noindent Flores, R. A., Primack J. R., Blumenthal G. R., Faber S. M., 1993, ApJ, 412, 443

\noindent Frenk, C. S., White, S. D. M., Efstathiou, G. P., Davis, M., 1985,
Nature, 317, 595

\noindent Frenk, C. S., White, S. D. M., Davis, M., Efstathiou, G. P., 1988,
ApJ, 327, 507

\noindent Gunn, J., Gott, J. R., 1972, ApJ, 209, 1

\noindent Hernquist, L., 1989 , Nature, Vol 340, 687

\noindent Hernquist, L., 1990, ApJ, 356, 359 (Hern)

\noindent Hoffman, Y., 1988, ApJ, 328, 489

\noindent Hoffman, Y., Shaham, J., 1985, ApJ, 297, 16

\noindent Jaffe, W., 1983, MNRAS, 202, 995


\noindent King, I. R., 1996, AJ, 71, 65

\noindent Mihos, J. C., Hernquist L., 1994, ApJ, 437, L47

\noindent Moutarde, F., Alimi, J.-M, Bouchet, F. R., Pellat, R., 1995, ApJ, 441, 10

\noindent Navarro, J. F., White, S. D., 1993, MNRAS, 265, 271

\noindent Navarro, J. F., Frenk, C., White , S. D. M., 1995, MNRAS, 275, 56 

\noindent Navarro, J. F., Frenk, C., White , S. D. M., 1995, MNRAS, 275, 720 (NFW)

\noindent Navarro, J. F., Frenk, C., White , S. D. M., 1996a, ApJ, 462, 563

\noindent Navarro, J. F., Frenk, C., White , S. D. M., 1996b, ApJ, in press.

\noindent Navarro J. F., Steinmetz M., 1996, ApJ, 438, 13

\noindent Persic M., Salucci P., Stel F., 1996, MNRAS, 281, 27 (PSS)

\noindent Quinn, P. J., Salmon, J. K., Zurek, W. H., 1986, Nature, 322, 329

\noindent Steinmetz, M., White, S. D. M., 1997, MNRAS, 289, 545

\noindent Teyssier, R., Chieze, J., Alimi, J.-M, 1997, ApJ, 480, 36

\noindent Thomas, P. A., Couchman, H. M. P., 1992, MNRAS, 257, 11

\noindent Tissera, P. B., Lambas, D. G., Abadi M., G., 1997, MNRAS, 286, 384

\noindent Tissera, P. B., Dom\ci nguez-Tenreiro, R. , Goldschmidt, P. , 1996,
Starburst Activity in Galaxies, Ed. J. Franco, R. Terlevich, A. Serrano, RevMexAA
(Conf Series), vol. 6, p 225

\noindent Tormen G., Bouchet F. R., White S. D. M., 1997, MNRAS, 286, 865

\noindent Warren, W. S.,  Quinn, P. J., Salmon, J. K., Zurek, W. H.,
1992, ApJ, 399, 405

\noindent White, S. D. M., Frenk, C., 1991, ApJ, 379, 52

\noindent Zurek, W. H., Salmon, J. K.,  Quinn, P. J., 1988, ApJ, 330, 519

\newpage
\section*{Table Captions}

Table I: Main parameters of simulations:
$N_{part}$ is the total number of particles simulated,
$\Omega_{b}$ is the baryonic density parameter, $b$ is the  bias parameter,
$\eta$ and $\rho_{star}$ $(\rm{gr/cm^{3}})$ are parameters of the star formation algorithm
(see text for details).\\

Table II: Main parameters of halos in the AP3M (sets I.1, II.1)
and SPH (sets I.2, I.3, II.2) runs: $N_{dark}$, $N_{star}$ and
$N_{gas}$ are the dark matter, star and gas numbers of particles, repectively, within the virial radius
used to resolved the halos. $M_{200}$ is the virial mass in units of $10^{12}M_{\odot}$, $r_{200}$ is the virial radius (kpc) and $r_{cross}$ the 
crossing radius as defined in section 3.5 (kpc). $c_{NFW}$ and $c_{TD}$
are the concentration parameters for the fits of NFW profile and
equation 6, respectively.
 GLO
 gives a general description
of the shape of the main baryonic object in the halo: spheroidal (S), disk (D)
(P denotes a pair of galactic-like objects).\\


\newpage
\section*{Figure Captions:}

Figure 1: Distribution of axis ratios $b/a$ and $c/a$ for halos
in the purely dynamical (solid triangles) and hydrodynamical (open
pentagons) simulations measured at the virial radius.\\

Figure 2: a) Semiaxes  $b/a$ (open) and $c/a$ (solid)
for 10 ellipsoids containing
from $10\%$ to $100\%$ of the dark mass within $r=100$ kpc of halo 1 (triangles)
and for halo 3 (pentagons) in simulation I.2; b) semiaxes $b/a$ (open) and $c/a$
(solid) for a halo in simulation I.2 (triangles) and I.3 (pentagons).
The values at the virial radii have also been plotted.\\

Figure 3: a) $\beta_{dark}$ as function of the radius for two halos in
the SPH (heavy lines) and N-body  (light lines) runs superposed;
b) total velocity dispersion $\sigma$
for 4 objects in simulation I.2 (heavy lines)
and for their corresponding counterparts in
simulation I.1 (light lines).\\

Figure 4: Dark matter density profiles for  4 halos in the
purely dynamical simulation (I.1, solid line)
and the best-fit given by NFW profile (dotted line).
The arrow shows the gravitational softening.\\

Figure 5: Dark matter density profiles for 2 halos in simulation I.1
(dotted lines) and their counterparts in simulation I.2 (solid lines).
The arrow shows the gravitational softening.\\

Figure 6: Dark matter density profiles for four halos
in the hydrodynamical run I.2 (solid line),
the best-fits given by NFW profile (dot-long dashed lines)
and by equation 6 (dotted line).
The arrow shows the gravitational softening.\\

Figure 7: The logarithm of the concentration parameters $c_{NFW}$ for halos
in I.1 (open pentagons), and $c_{TD}$ for halos in
 I.2 (solid squares) and I.3 (open triangles).
The curve is the predicted concentration from the collapse time
of the halos in purely dynamical models using NFW profiles.
The zero point of the relation for I.2 and I.3 has been
redefined to match the value for I.1 in order to be able to compare the
slopes. \\

Figure 8: The ratio of  the dark matter masses within $r=30$ kpc 
for halos in the hydrodynamical ($M_{dark}^{sph}$) simulations
to their counterparts in purely gravitational
($M_{dark}^{dyn}$) runs, versus their virial mass  $M_{200}$,
for halos in simulations I.2 (solid squares), I.3 (open triangles)
 and II.2 (open circles).\\

Figure 9: Density profiles of Halo 1 and Halo 4 in simulations I.2 and I.3:
dark matter density  profiles (solid lines) and baryonic density profiles
in I.2 (dotted lines) and I.3 (dot-long dashed lines).\\

Figure 10: Circular velocity curves  of the dark matter component
for three  halos in  simulation I.2. We also include the analytical fit given by Eq.(8) (solid line). The arrows indicate twice the softening length
used.\\

Figure 11 : Rotational velocity distributions for the baryonic particles of
objects in simulation
I.2 (solid triangles), circular velocity curves (according to equation 9) for
the total mass distribution (solid line), the dark matter only
(dot-dashed line)
and the baryonic component (dotted line).\\

Figure 12: Circular velocity curves for objects in simulation I.3 for
the total mass distribution (solid line), the dark matter only (dot-dashed line)
and the baryonic component (dotted line).
\\

Figure 13:
a) The virial  circular velocity ($V_{200}$) and, b) the maximum of 
the circular velocity curve ($V_{cir}^{max}$)
versus the crossing radius ($r_{cross}$) in units of the 
virial radius ($r_{200}$), for halos in  simulation I.2 (solid squares)
and simulation I.3 (open triangles)\\

Figure 14: The ratio of the baryonic mass inside $r_{cross}$ to 
the total baryonic
mass at $r_{200}$, versus the total virial mass,
for halos in simulations I.2 (solid squares), I.3 (open triangles) 
and II.2 (open circles).
We have omitted Halo 7 in set I since it hosts a pair of disk-like objects.
\\

Figure 15: Rotational velocity  for baryonic particles in four disk-like objects in I.2
(solid triangles) and the best fit given by the URC proposed by
Persic \eti (1996) (solid line).\\

\end{document}